\documentclass[twocolumn,apl,amsmath,amssymb,showpacs,superscriptaddress]{revtex4-2}
\usepackage{epsf}
\usepackage[utf8]{inputenc}
\usepackage{amsmath}
\usepackage{amsfonts}
\usepackage{amssymb}
\usepackage{makeidx}
\usepackage{color}
\usepackage{gensymb}
\usepackage{graphicx}
\usepackage{lipsum}
\usepackage{booktabs}
\usepackage{multirow}

\begin{document}

\title{Temperature dependent conductivity, dielectric relaxation, electrical modulus and impedance spectroscopy of  Ni substituted Na$_{3+2x}$Zr$_{2-x}$Ni$_{x}$Si$_2$PO$_{\rm 12}$}

\author{Ramcharan Meena}
\affiliation{Department of Physics, Indian Institute of Technology Delhi, Hauz Khas, New Delhi-110016, India}
\affiliation{Material Science Division, Inter-University Accelerator Center, Aruna Asaf Ali Road, New Delhi-110067, India}
\author{Rajendra S. Dhaka}
\email{rsdhaka@physics.iitd.ac.in}
\affiliation{Department of Physics, Indian Institute of Technology Delhi, Hauz Khas, New Delhi-110016, India}

\date{\today}      

\begin{abstract}

We investigate the structural, dielectric relaxation, electric modulus and impedance behavior of Ni-doped NASICON ceramic Na$_{3+2x}$Zr$_{2-x}$Ni$_{x}$Si$_2$PO$_{\rm 12}$ ($x=$ 0.05--0.2) prepared using the solid-state reaction method. The increase in dielectric constant with temperature and decrease with frequency is explained on the basis of space charge polarization using the two-layer model of Maxwell-Wagner relaxation. The dielectric loss peak at lower temperatures follows the Arrhenius-type behavior with frequency having activation energy of 0.27$\pm$0.01~eV of dipolar relaxation, suggests similar type of defects are responsible for all the doped samples. The real ($\epsilon$ $^{'}$) and imaginary  ($\epsilon$ $^{''}$) permittivity variation with frequency shows the broad relaxation behavior indicates the non-Debye type of relaxation in the measured temperature range. The permittivity values decrease with the amount of doping due to the increased number of charge carriers upon Ni doping at the Zr site. The impedance variation with frequency at various temperatures shows two types of relaxation for all the samples correspond to grain and grain boundary contribution in total impedance. The grain contributions are observed at higher frequencies, while grain-boundary contributions occur at the lower side of frequencies. The imaginary part of the electric modulus also shows two types of relaxation peaks for all the samples indicating similar activation energy at low temperatures and variable activation energy at higher temperatures. The fitting of the imaginary modulus using  KWW function shows the non-Debye type of relaxation. We find that all modulus curves merge with each other at low temperatures showing a similar type of relaxation, while curves at high temperatures show the dispersed behavior above the peak frequency. The {\it a.c.} conductivity data are fitted using the double power law confirming the grain and grain boundary contributions in total conductivity. The double power law exponent variation with temperature shows the small polaron hopping conduction over the measured temperature range for all the doped samples.            
    
\end{abstract}

\maketitle
\section{\noindent ~Introduction}

The NASICON structures [Sodium(\textbf{Na}) \textbf{S}uper \textbf{I}onic \textbf{CON}ductors] having the chemical composition of Na$_{1+x}$Zr$_2$Si$_x$P$_{3-x}$O$_{12}$ ($0\textless$x$\textless$3) have generated great research interest in the field of sodium-based solid-state batteries \cite{Wang_NS_22, Ruan_CI_19, Jiang_CEJ_23, Singh_Ionics_22, Li_IM_22}. These materials are considered excellent solid electrolyte because of zero leakage, non-flammability, suppressed dendrite penetration, low thermal expansion and their high stability against air as well as moisture. The Na$_{1+x}$Zr$_2$Si$_x$P$_{3-x}$O$_{12}$ ($0\textless$$x$$\textless$3), proposed by Hong and Goodenough in 1976, shows wide three-dimensional transport path for carriers to provide fast ion conduction, high structural tolerance, high hardness, high ionic transference number, wide electrochemical stability, high thermal stability, good chemical compatibility with electrode materials, high abundance and low-cost \cite{Goodenough_MRB_76, SapraWEE21, DwivediACSAEM21, PatiJMCA22, Wang_ESM_23, Oh_AMI_19, Li_IM_22, Lu_AEM_19, Sun_AFM_21}. The crystal structure varies with $x$, i.e., a monoclinic structure for $1.8\textless$$x$$\textless$2.2 with space group C2/c and rhombohedral phase with space group R$\bar{3}$c for other values of $x$. The monoclinic phase is stable at room temperature, and the contribution of the rhombohedral phase increases with an increase in temperature with a complete transformation at 150$\degree$C, which found to be due to shear deformation of the unit cell \cite{Lu_AEM_19, Sun_AFM_21,  Oh_AMI_19}. Here, both (monoclinic and rhombohedral) structure contains SiO$_4$/PO$_4$ tetrahedra that are linked to ZrO$_6$ octahedra through corner shared oxygen ions. The ionic conductivity in NASICON samples is due to the Na ion migration through bottleneck triangles formed by O atoms of two polyhedrons \cite{Sun_AFM_21,  Oh_AMI_19, Goodenough_MRB_76}. The structural arrangements show two types of Na$_1$ and Na$_2$ ions in the rhombohedral phase has only transport path Na$_1$--Na$_2$, while monoclinic structure contains three types of Na ions, as Na$_2$ ion-sites further split into the Na$_2$ and Na$_3$, which provides additional transport paths (Na$_1$--Na$_2$ and Na$_1$--Na$_3$) and consequently the higher conductivity and lower activation energy. Out of these available Na sites, two-thirds of sites are occupied and only one-third are available for conduction. The Na ion conduction occurs through the channel made by SiO$_4$/PO$_4$ tetrahedra and ZrO$_6$ octahedra units. The highest conductivity is observed for the $x=$ 2, i.e., the Na$_3$Zr$_{2}$Si$_2$PO$_{\rm 12}$ composition; however, the total ionic conductivity is of the order of $10^{-4}$ S cm$^{-1}$ at room temperature, and it is still two orders less as compared to commercial liquid electrolytes ($10^{-2}$ S/cm), which puts a limit to use these NASICON materials in commercial solid-state batteries. At the same time, it is reported that the conductivity depends on the method and treatment during the sample preparation \cite{Sun_AFM_21, Goodenough_MRB_76, Jian_AM_17}. 

 The NASICON ceramic sample preparation requires high sintering temperature, which may create a composition imbalance in the system due to the volatile nature of Na and P ions, resulting in an unwanted impurity phase in the sample. These impurities accumulated at the grain boundaries give a reduction in overall conductivity. The required stoichiometry of Na and P can be maintained by lowering the sintering temperature at the cost of decreased density and dense microstructure. In order to reduce the overall sintering temperature and increase in total conductivity, various methods such as liquid phase sintering \cite{Wang_NS_22, Oh_AMI_19}, ultra-fast sintering \cite{Jiang_CEJ_23}, rapid microwave sintering \cite{Wang_JPS_21}, doping of multi valent atoms \cite{Wang_ESM_23, Sun_AFM_21, Lu_AEM_19, Ma_CM_16}, co-doping of different valence elements \cite{Yang_AEL_20, Thirupathi_JPCC_21, Ran_ESM_21}, varying the weight ratio of initial precursors \cite{Ruan_CI_19} and various other methods of sample preparation are used in the past \cite{Ma_CM_16, Dubey_AEM_21}. In liquid phase sintering various compounds like Na$_2$SiO$_3$ \cite{Wang_NS_22, Oh_AMI_19}, Na$_3$BO$_3$ \cite{Noi_JACS_18, Suzuki_SM_18} having the lower melting point as compared to the NASICON compound are added to the host matrix, which reduces the overall sintering temperature and helps to enhance the conductivity and density similar to the samples prepared using high-temperature sintering technique \cite{SapraWEE21, DwivediACSAEM21, PatiJMCA22}. In the liquid phase sintering additives diffuse inside the host matrix and lead to an enlarged bottleneck area for Na-ion conduction giving an overall increase in conductivity with reduced activation energy \cite{Wang_NS_22, Oh_AMI_19, Noi_JACS_18, Suzuki_SM_18}. Similarly, ultra-fast sintering and rapid microwave sintering give the density and conductivity similar or higher to the sample prepared using ceramic sintering and reduce the volatilization of sodium \cite{Jiang_CEJ_23, Wang_JPS_21}. 
 
 In general, the conductivity of NASICON ceramics is altered by doping at the Zr site, as the ionic radius mismatch leads to changes in the bottleneck area available for conduction, reduces the densified temperature, and optimizes the grain-boundary phase with increased conductivity. The maximum ionic conductivity for doped samples depends on the valence state of the dopant atom, concentration, and ionic radius of the doped element. For example, the doping of divalent and/or trivalent atoms at the Zr site increases the density and lowers the monoclinic to rhombohedral phase transition due to the shear deformation of the unit cell, as well as increases the number of free carriers leading to an increase in total conductivity.\cite{Chen_JALCOM_2018, Khakpour_ECA_16, Jolley_JASC_15, Wang_ESM_23, Lu_AEM_19, Pal_JPC_20, Kumar_CPL_21, Park_JPS_18}. Also, the doping of tetravalent/pentavalent atoms having an ionic radius larger than Zr gives the enlargement in bottleneck area for Na-ion diffusion exhibits the increase in conductivity \cite{Khakpour_ECA_16, Liu_API_20}. The effect of excess Na and P has also been studied in the past to improve the conductivity of NASICON materials as it compensates for the loss of Na and P during sintering, where it was found that samples with excess sodium lower the required sintering temperature and exhibit the highest conductivity \cite{Naqash_SSI_19, Narayanan_SSI_19, Park_AMI_16}. 

It has been reported that NASICON doped with divalent elements at the Zr site can alter its lattice constants, free carrier concentration, and crystal structure, enhancing the total conductivity as well as improves the density of these ceramics \cite{Lu_AEM_19, Chen_JECS_18, Jolley_Ionics_15}. The divalent dopant creates more interstitial charge carriers to compensate for the charge imbalance in the system and regulates the grain boundary phase leading to increases the conductivity. The size of the dopant having an ionic radius larger than Zr ion shows an increase in the bottleneck area and reduces the migration barrier between the hopping sites available for Na-ion migration \cite{Chen_JECS_18, Li_IM_22}. The NASICON based materials having a wide 3-dimensional network, easy preparation, and high ionic conductivity make as suitable candidates for various fields like solid-state Na$^+$ ion--batteries, gas sensing, super-capacitors, ion-selective electrode, and microwave absorption, etc. \cite{Li_IM_22, Rao_SSI_21, Singh_JES_21, Ma_JMCA_19, Chen_JECS_18}. In some cases, the lower density of NASICON makes it a good choice for the above applications due to its reduced weight \cite{Chen_JECS_18}. However, there are very few studies available in the literature on microwave absorption and dielectric properties of NASICON materials \cite{Chen_JECS_18, Dubey_AEM_21, Chen_ML_18}, but limited at room temperature or high temperature only. In our recent study, we have investigated the dielectric properties of NASICON materials and its relaxation mechanism at lower temperatures \cite{Meena_CI_22}. However, to the best of our knowledge, there are no reports available in the literature on dielectric properties of divalent doped NASICON materials at lower temperatures; which are vital and motivated us to study the Ni-doped Na$_3$Zr$_2$Si$_2$PO$_{12}$ to make these materials useful for practical applications. 

Therefore, for this paper, we have synthesized NASICON structured bulk samples having the chemical composition of Na$_{3+2x}$Zr$_{2-x}$Ni$_{x}$Si$_2$PO$_{\rm 12}$ ($x=$ 0.05--0.2) using the solid-state reaction method and investigated its structural and electrical transport properties. The phase and purity are checked using XRD measurement at room temperature. The structural analysis using the Rietveld refinement confirms the monoclinic phase (space group-C 2/c) for all samples. The temperature-dependent resistivity measurement shows the semiconducting behavior. Moreover, the {\it a.c.} and {\it d.c.} conductivity are measured and data are fitted using the appropriate models. Further, the electric permittivity, dielectric loss, total impedance, and phase angle are studied as a function of frequency at different temperatures. The relevant models are used to understand the relaxation dynamic. 

\section{\noindent ~Experimental}

 The polycrystalline bulk Na$_{3+2x}$Zr$_{2-x}$Ni$_{x}$Si$_2$PO$_{\rm 12}$ ($x=$ 0.05--0.2) samples are prepared using the conventional solid-state reaction method. The SiO$_2$ crystals were preheated at 160$^\circ$C for 16 hrs to remove any moisture present in silica. All the precursors such as Na$_2$CO$_3$ (purity 99.5\%), ZrO$_2$ (purity 99.5\%), NiO(purity 99.8\%), SiO$_2$ (purity 99\%) and NH$_4$H$_2$PO$_4$ (purity 99.9\%) were taken in stoichiometric ratio with 10\% excess of Na$_2$CO$_3$ and NH$_4$H$_2$PO$_4$ to compensate the loss of Na and P ions due to high temperature sintering. All of these precursors were thoroughly ground using an agate mortar pestle for uniform mixing of all chemicals and the initial calcination reaction was performed at 1000$^\circ$C for 6 hrs in air. The resultant powder became light green in color and were obtained in colloidal form. The calcinated powder was further grinded using an agate mortar pestle, to have a uniform distribution of particles. The grinded powder was pressed into circular pellets having a diameter of 10~mm with a thickness equal to 1.8~ mm using the hydraulic pressure of 1000~psi with a holding time of 5 min. Finally, these prepared pellets were sintered at 1150$^\circ$C for 15 hrs with heating and cooling rates of $5^\circ$C/min.
 
	The structural investigations of all the samples (at room temperature) were performed by the x-ray diffraction (XRD) using the Bruker D8 Advance diffractometer in Bragg-Brentano geometry equipped with the Cu-K$_\alpha$ radiation of wave-length ($\lambda$=1.54~$A^\circ$) in the 2$\theta$ range of 10$^{\circ}$- 80$^{\circ}$ with an accelerating voltage of 40~V and 30~mA current. The XRD patterens were analyzed using the FullProf Suite software where various parameters like scale factor, unit cell parameters, structure factor, position parameters, and occupancy factors were refined. The surface morphology of the Ni-doped samples was obtained using JEOL JSM-7610F Schottky Field Emission Scanning Electron Microscope. As these samples are in insulating nature, Pt metal thin film of thickness around 10 nm was coated on all samples to avoid any charging effect. The temperature-dependent {\it d.c.} resistivity ($\rho$-$T$) data were recorded using Keithley make electrometer (Model-6517B) by applying the source voltage of 1~V and measuring the resultant current. 
	
	The frequency-dependent measurements are performed in 4~T (terminal) configuration using an Agilent LCR meter (Model-E4980A) in parallel circuit mode. To avoid any error due to parasitic residual components open, short, and cable length corrections (calibrations) were performed before starting any frequency-dependent measurements. Both sides of these pellets were coated using silver paste and dried at 150 $^\circ$C for 1 hr before performing the temperature-dependent (100--400~K) measurements like parallel capacitance ($C_P$), dielectric loss (D=tan$\delta$), total impedance ($\lvert Z \rvert$) and phase angle ($\theta$) parameters in the frequency range of 20~Hz-2~MHz.  During all of these measurements, an input perturbation of 1~Volt {\it a.c.} signal was applied and the resultant current was measured. The temperature during these measurements was recorded using a Pt-100 sensor and the temperature was varied using the Lakeshore temperature controller (Model-340). The experimental data are recorded keeping the temperature stability of 100~mK with a stability time of 1 minute during entire measurements. The electrometer, LCR meter, and temperature controller are interfaced with LabView software for automatic measurements. The sample holder was placed in a liquid nitrogen-based dipstick bath for uniform cooling across the samples. The vacuum of the order of 10$^{-3}$ mbar was maintained near the sample during the measurements. 
	
\section{\noindent ~Results and discussion}

\begin{figure*}
\centering
\includegraphics[width=1.0\textwidth]{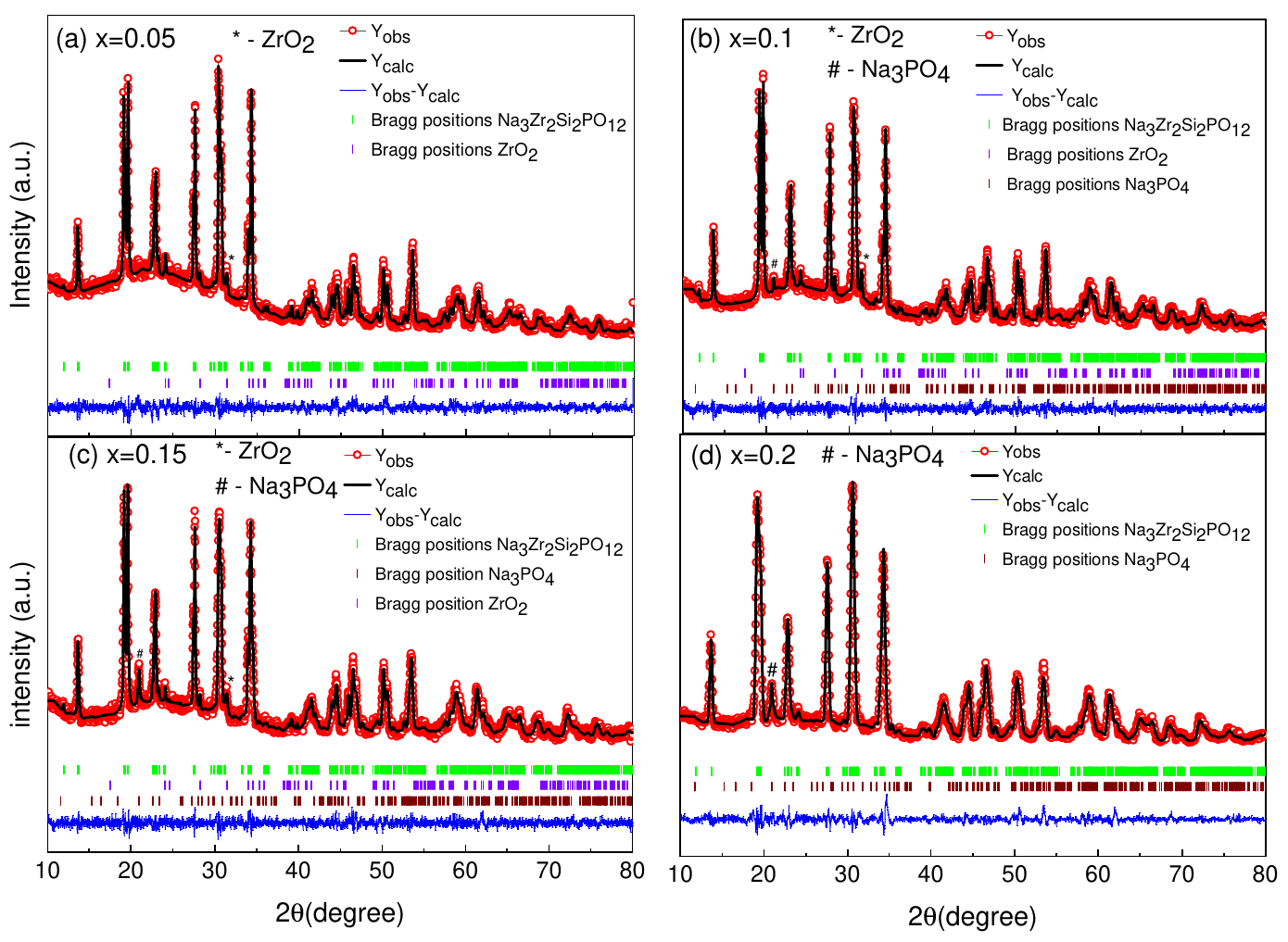}
\caption {(a--d) The Rietveld refined room temperature x-ray diffraction (XRD) patterns of (a) Na$_{3.1}$Zr$_{1.95}$Ni$_{0.05}$Si$_2$PO$_{\rm 12}$, (b) Na$_{3.2}$Zr$_{1.9}$Ni$_{0.1}$Si$_2$PO$_{\rm 12}$, (c) Na$_{3.3}$Zr$_{1.85}$Ni$_{0.15}$Si$_2$PO$_{\rm 12}$, and (d) Na$_{3.4}$Zr$_{1.8}$Ni$_{0.2}$Si$_2$PO$_{\rm 12}$ samples. The observed, calculated, and the difference between observed and calculated patterns are shown by the open red circle, black solid line, and continuous blue line, respectively;  green, violet and brown vertical markers represent the Bragg positions corresponding of the C 2/c,  P 1 21/c and P -4 21 c space groups, respectively.}.
 \label{XRD}
\end{figure*}

Figures~1(a--d) shows the Rietveld refined room temperature XRD patterns of the Na$_{3+2x}$Zr$_{2-x}$Ni$_{x}$Si$_2$PO$_{\rm 12}$ ($x=$ 0.05--0.2) samples where the brackets in each pattern show the corresponding Ni doping concentration in each panel. The XRD patterns are analyzed using the Rietveld refinement method by taking the pseudo-voigt peak shape and linear interpolation between the data points in Fullprof Suite software. The initial values of crystal structure, space group, unit cell parameters, and atomic coordinates were taken from the matching references from the crystallographic database. It is found that the observed patterns and calculated patterns overlap with each other indicating the required phase formation. The goodness of fit (reduced chi-square $\chi$$^2$ value) is found to be between 1.8--2.2 confirming the monoclinic phase having C~2/c space group for all the samples. It is also observed that a small amount of ZrO$_2$ (denoted by *) and Na$_3$PO$_4$ (denoted by $\#$ ) impurity phases are present with the main NASICON phase. The intensity of the ZrO$_2$ phase is decreased with an increase in doping and the opposite is found for the Na$_3$PO$_4$ phase. This type of impurity phases were also reported in literature for the NASICON samples \cite{Jolley_Ionics_15}. The experimentally obtained lattice parameters after fitting are shown in table \ref{tab:table1}. It is clear from the table the values of $a$ and $c$ remain nearly constant, while $b$ and $\beta$ increase with Ni doping; however, the total volume initially decreased and then increased with further increase in the doping level, as the number of carriers increases with doping may increase the unit cell volume. The obtained lattice parameters are in close matching with the values reported in literature \cite{Lu_AEM_19, Jolley_JASC_15}. 

\begin{table}[htbp]
\caption{The room temperature structural parameters of Ni-doped NASICON samples obtained from profile fittings using Rietveld refinement method.}
\label{tab:table1}
\begin{center}
\begin{tabular}{|c|c|c|c|c|c|c|}
\hline
\text{sample details}  & \multicolumn{ 4}{c|}{\text{lattice parameters}}  & \text{volume} & \text{$\chi$$^2$} \\ \hline
 &  a (A$^\circ$) & b (A$^\circ$) & c (A$^\circ$) & $\beta$$^{\circ}$   & (A$^\circ$)$^3$  &  \\ \hline
$x=0.05$ & 15.65  & 9.05 & 9.21 & 123.81  &  1085.45 & 1.82 \\ \hline
$x=0.1$ & 15.65  & 9.05 & 9.20 & 123.88  &  1081.62 & 2.26 \\ \hline
$x=0.15$ & 15.66  & 9.06 & 9.20 & 123.95  &  1082.55 & 1.84 \\ \hline
$x=0.2$ & 15.65  & 9.08 & 9.21 & 124.02  &  1083.36 & 1.89 \\ \hline
\end{tabular}
\end{center}
\end{table}

\begin{figure*} 
\includegraphics[width=1\textwidth]{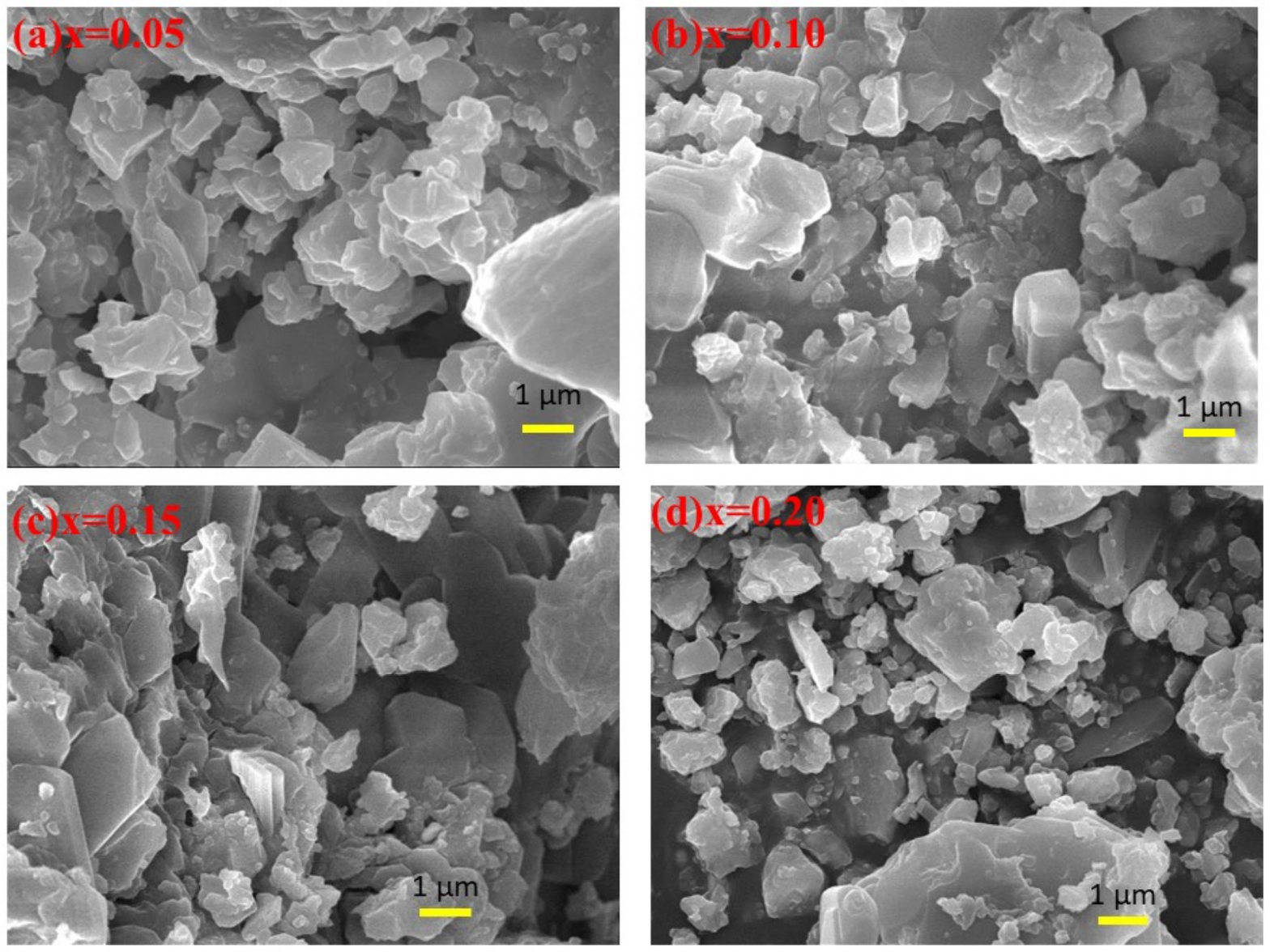}
\caption {The scanning electron microscope (SEM) images (a--d) of Na$_{3+2x}$Zr$_{2-x}$Ni$_{x}$Si$_2$PO$_{\rm 12}$ $(x=0.05-0.2)$ samples showing the non-uniform distribution of particles. All the images are taken at a similar scale shown at the bottom of each panel. } 
\label{SEM}
\end{figure*}

\begin{figure} 
\includegraphics[width=3.4in]{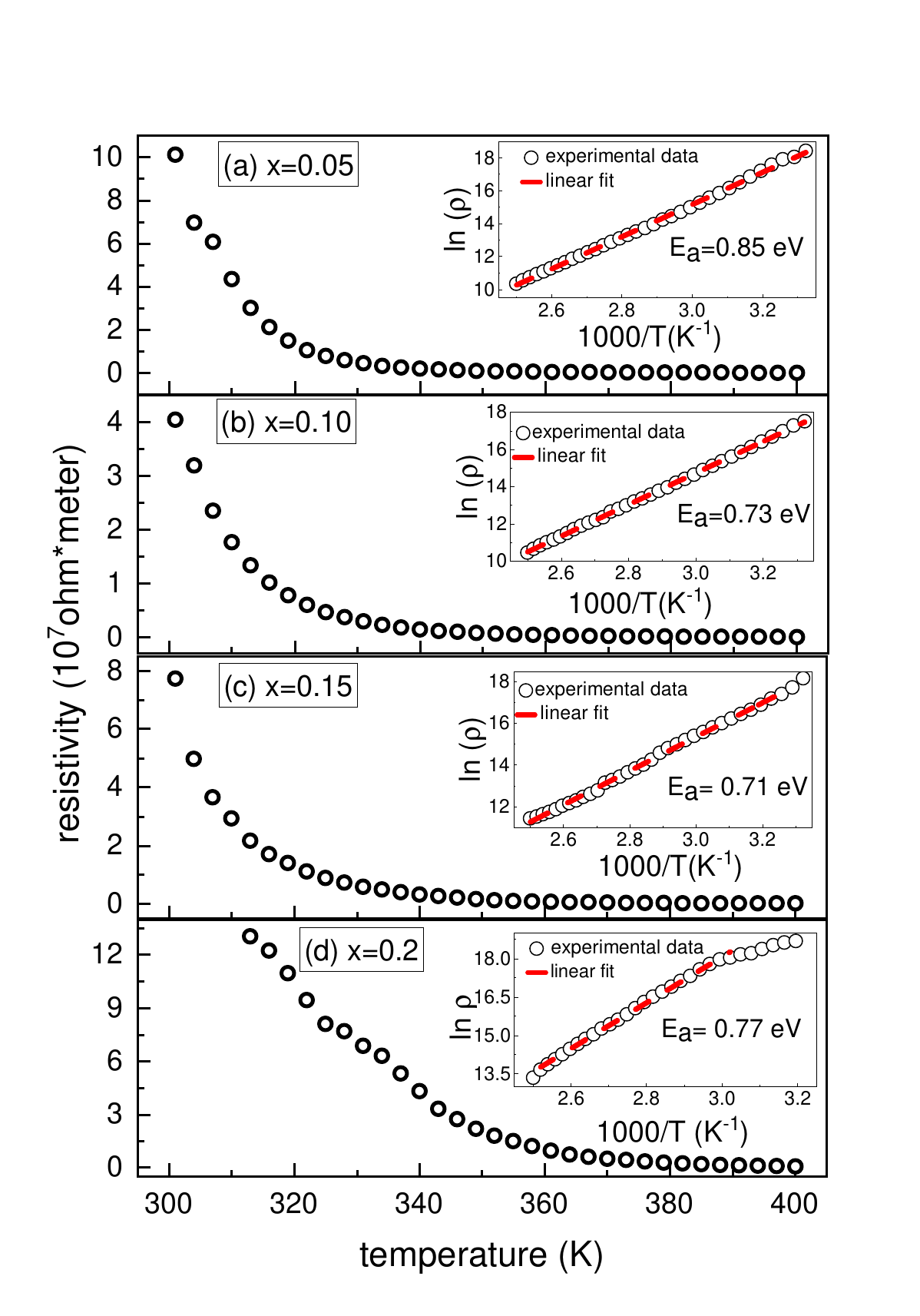}
\caption {The temperature dependent resistivity variation of the Na$_{3+2x}$Zr$_{2-x}$Ni$_{x}$Si$_2$PO$_{\rm 12}$ $(x=0.05-0.2)$ samples in panels (a--d). The inset in each panel shows the Arrhenius thermal conduction using the plot of (ln~$\rho$ versus 1000/T) to find the activation energy of thermal conduction. Here, the open circles represent the experimental data and the solid line represents the linear fit.} 
\label{RT}
\end{figure}

The influence of doping on the surface morphology of Na$_{3+2x}$Zr$_{2-x}$Ni$_{x}$Si$_2$PO$_{\rm 12}$ ($x=$ 0.05--0.2) samples are presented in Figures~\ref{SEM}(a--d). All these samples show a similar morphology having a non-uniform distribution of grains with dense microstructure. The smaller particles show the tendency to agglomerate into bigger particles with negligible porosity are found for all the samples. It is observed that the particle size decreases with the increase of the doping concentration. The temperature-dependent resistivity measurements are performed in the temperature range of 300--400~K in order to determine the type of conduction mechanism followed by the carriers and its associated activation energy. All the samples show the negative temperature coefficient of resistance, i.e., the resistivity increases with a decrease in temperature. The resistivity measurements could not performed below room temperature due to the highly insulating (resistivity increases abruptly with decreasing temperature) nature of all the samples. Figures~\ref{RT}(a--d) show the resistivity versus temperature behavior of Ni-doped NASICON samples. The carrier transport mechanism and its associated activation energy are determined using the Arrhenius thermal activation model; according to this model, in a high-temperature regime conduction takes place due to the activation of charge carriers between the valence band and conduction band. The activation energy of the charge carriers can be calculated using the Arrhenius equation given by \cite{Meena_CI_22}    
\begin{equation}
    \rho_{T}=\rho_{0}~exp(\frac{E_a}{{k_B} T})
\label{Arrhenius}
\end{equation}
Here, the $\rho_{T}$ is the experimentally measured resistivity at temperature T, $\rho_{0}$ is the pre-exponential factor of resistivity, $E_a$ is the associated activation energy of thermally generated charge carriers and $k_B$ is the Boltzmann constant. Using equation~\ref{Arrhenius}, the slope of ln($\rho_{T}$) versus  ($\frac{1000}{T}$) [shown in the insets of Figures~\ref{RT}(a--d)] gives the estimate of activation energies of charge carriers, which are found to be 0.85, 0.73, 0.71 and 0.77 eV for the $x=$ 0.05, 0.1, 0.15 and 0.2 samples, respectively. It is observed that the activation energy E$_a$ is initially decreasing due to the increased number of charge carriers with Ni doping (as two Na ions are added for Ni ion to compensate the charge misbalance in the samples) and increased with further doping due to the presence of an impurity phase (ZrO$_2$ and Na$_3$PO$_4$) and increased scattering between the charge carriers. 

\begin{figure*}
    \centering
    \includegraphics[width=1.0\textwidth]{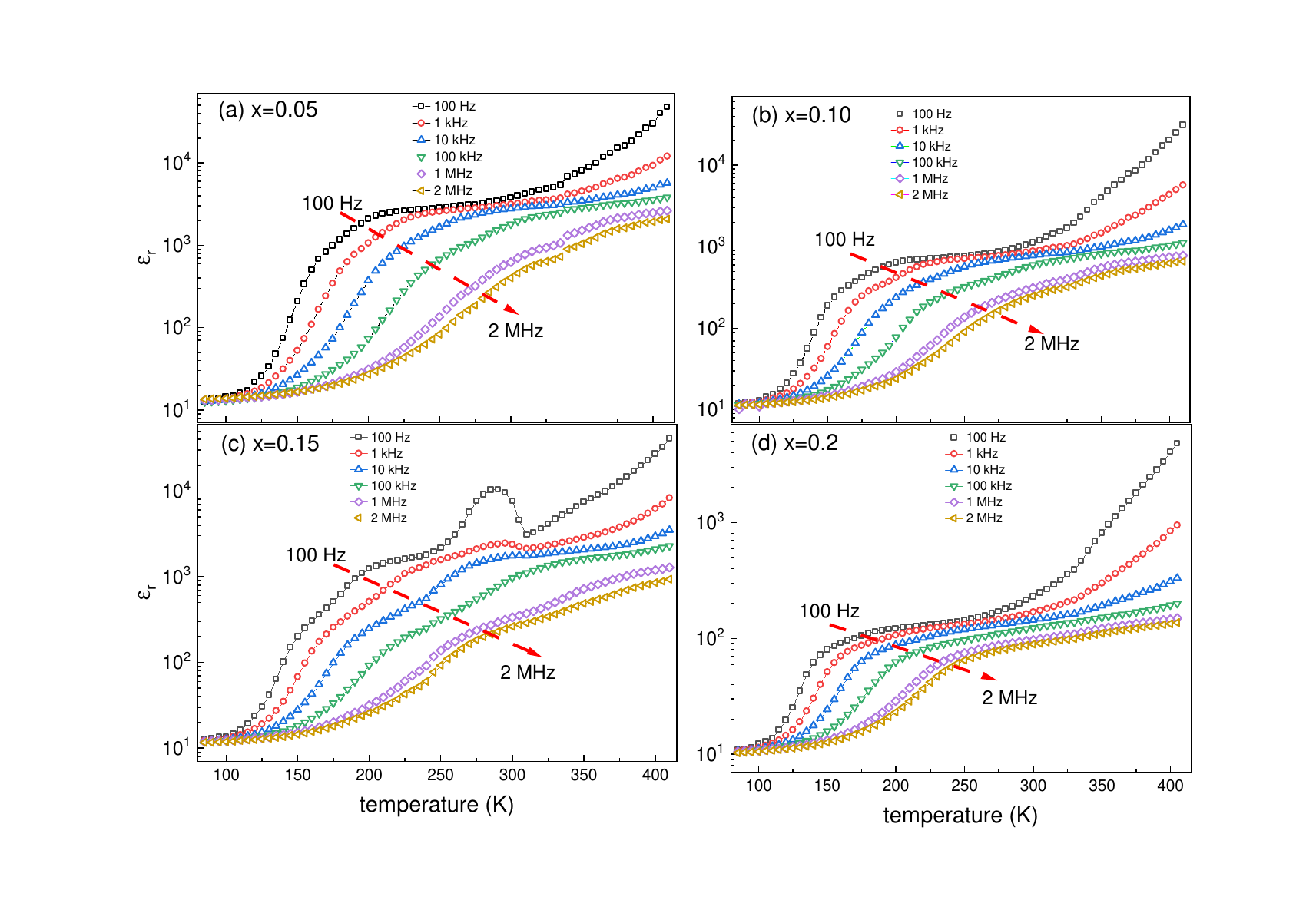}
    \caption{The temperature dependent dielectric constant variation of the Na$_{3+2x}$Zr$_{2-x}$Ni$_{x}$Si$_2$PO$_{\rm 12}$ ($x=$ 0.05--0.2) samples at different selected frequencies, as shown in panels (a--d). The arrows in each panel show the shifting of the relaxation peak toward high-temperature side with an increase in the frequency.}
    \label{CT}
\end{figure*}

Figures~\ref{CT}(a--d) show the temperature dependence of the dielectric constant (electric permittivity) at selected frequencies for the Na$_{3+2x}$Zr$_{2-x}$Ni$_{x}$Si$_2$PO$_{\rm 12}$ $(x=0.05-0.2)$ samples. The variation in dielectric constant shows low values of permittivity having the pleatu like structure at lower temperatures and increasing at higher temperatures. All the Ni-doped samples show the relaxation type behavior (as shown by the arrow) where the peak shifts towards the higher temperature side with an increase in the frequency. It is also found that the increase in dielectric constant is abrupt as we increase the Ni-doping. There are two important factors that contribute to the relaxation process of dielectric material, (i) the rate of polarization formed, and (ii) the frequency of the applied field. Once the sample temperature is high, the rate of polarization formation increases giving the relaxation occurs at a higher frequency \cite{Kumar_JALCOM_16, Lin_PRB_05}.  The dielectric constant values at a particular temperature can be calculated using the following relation:   
 \begin{equation}
    \epsilon_r(\omega)=\frac{C_P}{C_0}=\frac{C_P d}{\epsilon_0 A}
\end{equation}
here, the $\epsilon_r$($\omega$) is the dielectric constant measured at an angular frequency $\omega$, $C_P$ is the measured parallel mode capacitance, $C_0$ is vacuum capacitance (when no dielectric is present between parallel plates) given by $C_0$=$\epsilon_0$ $\frac{A}{d}$, where $\epsilon_0$ is the free space permittivity ($8.85 \times 10^{-12}$ F/m), $A$ is the cross-sectional area of electrodes and $d$ is the thickness of dielectric material.

\begin{figure*}
\centering
\includegraphics[width=1.0\textwidth]{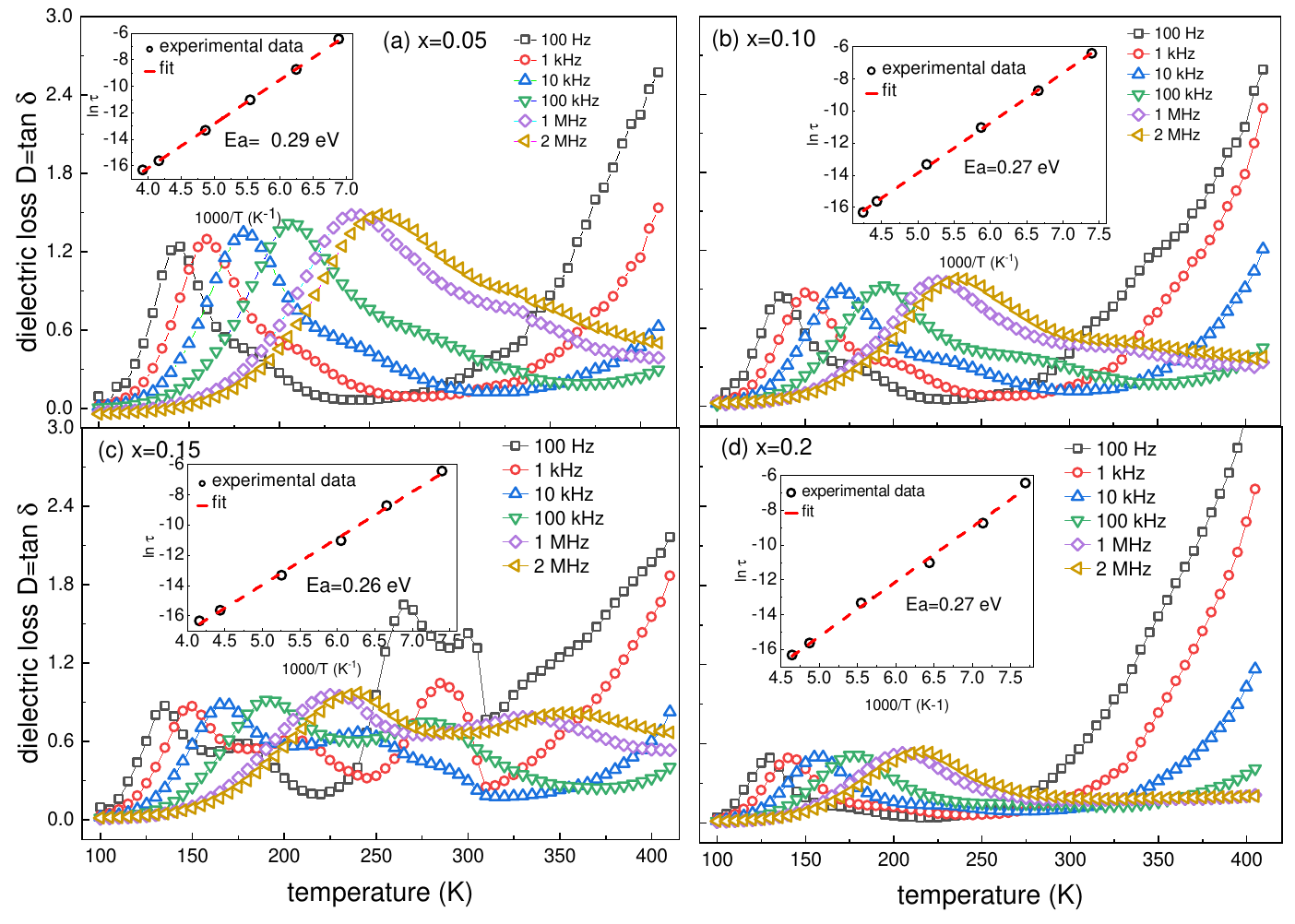}
\caption {The temperature dependent dielectric loss variation of Na$_{3+2x}$Zr$_{2-x}$Ni$_{x}$Si$_2$PO$_{\rm 12}$ $(x=0.05-0.2)$ samples at various selected frequencies, are shown in panels (a--d). The loss peak shifts towards the high-temperature side with an increase in the frequency. The inset in each panel shows the plot of (ln $\tau$ versus 1000/T), where the open circle represents the experimental data and the solid lines show the Arrhenius fit. } 
\label{DT}
\end{figure*}

  It is observed from experimental data that the electric permittivity depends on the applied frequency and the measured temperature. The increasing behavior of electric permittivity with temperature and decreasing with the frequency can be explained using the Maxwell-Wagner two-layer model of space charge polarization; according to this model, a material is considered to be made up of large well-conducting grains separated by small poorly conducting grain boundaries, and the charge carries (dipoles in this case) under the influence of applied field piles up at the grain-boundary creates the local interfacial polarization \cite{Lunkenheimer_PRB_04, Ke_Je_09}. The temperature dependence of electric permittivity can be explained as follows, as we increase the temperature, the bound charge carriers have sufficient thermal energy to orient themselves in the direction of the applied field, producing the larger values of dielectric constant at higher temperatures \cite{Lin_PRB_05}. The frequency dependence of the dielectric constant can be explained using the interfacial polarization model; according to this model, at lower values of frequency charge carriers move inside the grain and pile up at the grain boundary producing interfacial polarization inside the material, results in the higher values of dielectric constant; while at higher frequencies the field variation are so rapid that the carriers are no longer able to follow the applied field gives a reduction in total polarization results in the lower values of dielectric constant \cite{Ke_Je_09, Li_PRB_07}. The total contribution in dielectric constant at lower and higher frequencies is dominated by the grain boundary and grains, respectively \cite{Kumar_JALCOM_16, Lin_PRB_05}. 

Figures~\ref{DT}(a--d) show the dielectric loss data of Ni-doped NASICON samples at selected frequencies as a function of temperature. The dielectric constant variation with temperature (Figure~\ref{CT}) shows a sudden decrease in the dielectric constant at lower temperatures and appears as a peak in dielectric loss spectra. In other words, once the frequency of the applied electric field becomes equal to the hopping frequency of charge carriers results in a peak of loss spectra due to charge carrier localization. All the relaxation peaks follow the condition given by $\omega$$\tau$=1, where $\omega$=2$\pi$$f$ is the applied angular frequency over the carriers and $\tau$ is the relaxation time for the charge carriers. The defects and impurities inside the material act as a pinning center for the dipoles giving the lag of polarization as compared to the applied electric field producing a loss peak. We find that at lower temperatures the loss is nearly the same for all the selected frequencies while at high temperatures large dielectric loss is visible at lower values of frequency mainly due to the space-charge polarization and can be explained using the Schokely-Read mechanism \cite{Shockley_PR_1952}. According to this mechanism, the impurity and defects center captures the electron at the surface, generating surface-charge polarization. This electron capture process increases with temperature resulting in higher values of dielectric loss at high temperatures \cite{Yang_JAP_12}. It is observed that the relaxation peak broadens with an increase in frequency suggesting the spread of relaxation time, which can be explained using the relation $\omega$$\tau$=1. The relaxation peak shifts towards the high-temperature side with an increase in frequency, suggesting Arrhenius-type thermally activated relaxation for all the samples. 

The temperature dependence of relaxation time follows the relation given by expression as below \cite{Ang_PRB_00,  Shulman_JACS_00, Jia_JAP_11, Meena_CI_22}
\begin{equation}
     \tau = \tau_0 ~ exp(\frac{E_{relax}}{k_B T})
\label{tr}
 \end{equation}
Here, the $\tau_0$ is the characteristic relaxation time of charge carriers has the order of atom vibrational period ($10^{-13}$ sec), $E_{relax}$ is the activation energy of charge carrier relaxation, k$_B$ is the Boltzmann constant and T is the measured temperature. The activation energy of dipolar relaxation is determined using the slope of the linear plot between ln $\tau$ versus $(\frac{1000}{T})$, and the characteristic relaxation time using the intercept in the curve for each sample. The activation energy of all the Ni-doped samples is shown in the inset of Figures~\ref{DT}(a--d). The experimentally obtained values of activation energy and characteristic relaxation time are summarized in Table~\ref{tab:table2}. We find that all the samples have similar values of activation energy indicating a similar type of relaxation. The magnitude of activation energy indicates the polaron-type hopping conduction inside the material over the measured temperature range \cite{Dutta_MRB_11}. The characteristic relaxation time is found in agreement with the values reported in the literature \cite{Ang_PRB_00,  Shulman_JACS_00, Jia_JAP_11, Meena_CI_22}. 

\begin{table}[htbp]
\caption{The calculated activation energy of dielectric relaxation ($E_{relax}$) and characterstic releaxation time ($\tau_{0}$) for Na$_{3+2x}$Zr$_{2-x}$Ni$_{x}$Si$_2$PO$_{\rm 12}$ $(x=0.05-0.2)$ samples.}
\label{tab:table2}
\begin{center}
\begin{tabular}{|c|c|c|}
\hline
\text{sample details} & \text{activation energy}  & \text{relaxation time}  \\ \hline
& {$E_{relax}$} (eV) & $\tau_{0}$ * (10$^{-13}$ s)   \\ \hline
$x=0.05$ & 0.29  & 1.49   \\ \hline
$x=0.10$ & 0.27  & 1.52   \\ \hline
$x=0.15$ & 0.26  & 1.81   \\ \hline
$x=0.20$ & 0.27  & 0.61   \\ \hline
\end{tabular}
\end{center}
\end{table}

\begin{figure*}
    \centering
    \includegraphics[width=1.0\textwidth]{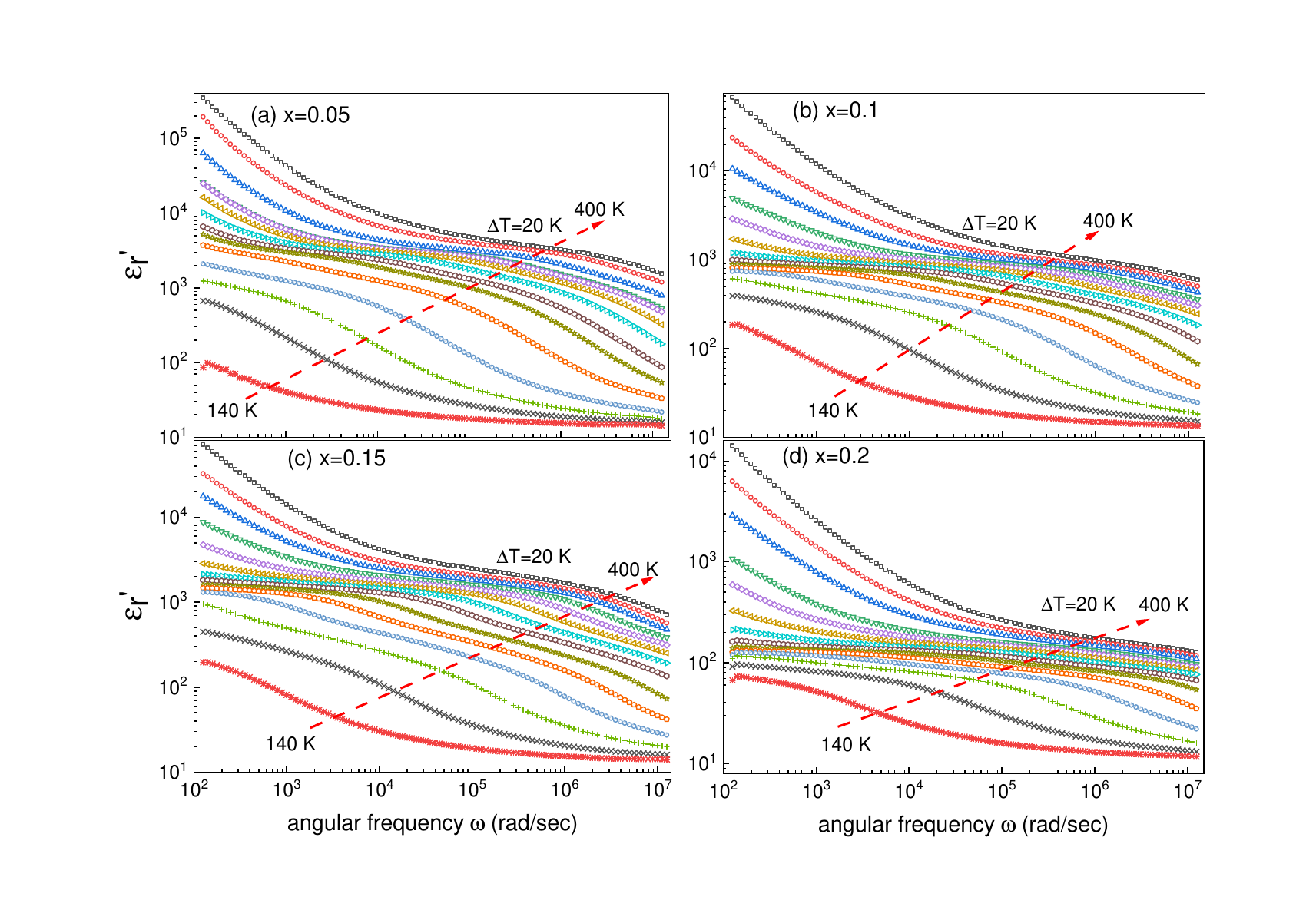}
    \caption{The temperature dependence of real part of electric permittivity of Na$_{3+2x}$Zr$_{2-x}$Ni$_{x}$Si$_2$PO$_{\rm 12}$ $(x=0.05-0.2)$  samples as a function of frequency, as shown in panels (a--d). The arrows indicate an increase in relaxation frequency towards the high-temperature side with an increase in temperature from 140--400~K.}
    \label{RE}
\end{figure*}

\begin{figure*} 
\centering
\includegraphics[width=1.0\textwidth]{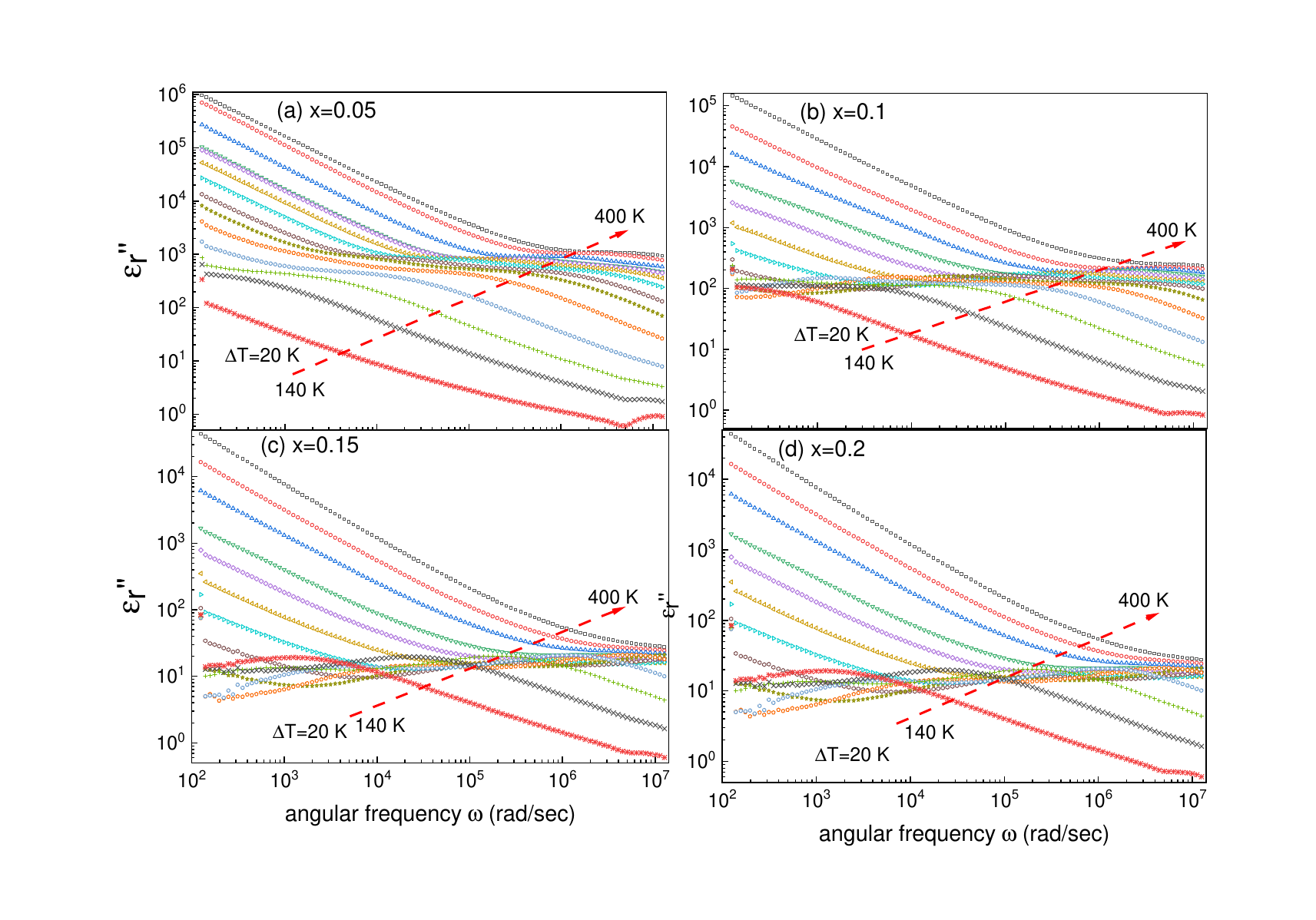}
\caption {Temperature dependence of imaginary part of electric permittivity Na$_{3+2x}$Zr$_{2-x}$Ni$_{x}$Si$_2$PO$_{\rm 12}$ $(x=0.05-0.2)$  samples as a function of frequency are shown in figure (a-d). The arrow shows an increase in relaxation frequency towards the high-temperature side with an increase in temperature from 140 K- 400 K.} 
\label{IE}
\end{figure*}

To further understand the relaxation phenomena, the frequency dependence of real and imaginary parts of the electrical permittivity at various temperatures are shown in Figures~\ref{RE} and \ref{IE}, respectively. The real ($\epsilon$ $^{'}$) and imaginary ($\epsilon$ $^{''}$) part of electrical permittivity represents the energy storage and loss of energy during each cycle of the applied electric field, respectively. It is found that the dielectric constant reduces with the Ni concentration, as the number of free carriers increase with the doping. These free-charge carriers decrease the amount of polarization produced by the applied electric field giving the lower values of dielectric constant with increase in the Ni doping. We observe that the dielectric constant or electrical permittivity increases with temperature and inverse behavior is followed with frequency for all the samples. All the Ni-doped samples show the strong dispersion behavior of electric permittivity, which is due to the thermally activated charge carriers like space charge carriers, defects, and their related complexes \cite{Li_JAP_14, Pu_JALCOM_16, Rehman_JAP_15}. The temperature dependence of ($\epsilon$ $^{''}$) versus frequency data are shown in Figures~\ref{IE}(a--d) showing the broad relaxation peak. These peaks shift towards the high-frequency side with an increase in temperature representing the thermally activated relaxation behavior. The shifting of peaks towards the higher frequency side with an increase in temperature is due to mobilizing the charge carriers and reduction in the pinning of defects, which leads to the large numbers of carriers in the relaxation process \cite{Dutta_MRB_11, Rehman_JAP_15}. The high temperature leads to a higher rate of polarization due to the large thermal energy available with carriers; resulting in the relaxation occurring at the higher frequency side. The large values of ($\epsilon$ $^{''}$) at lower frequency values are due to the contribution from {\it d.c.} conductivity \cite{Pu_JALCOM_16, Dutta_MRB_11}. The frequency-dependent behavior of loss peak (shown in Figure~\ref{IE}) shows the broad relaxation peaks indicate the distribution of relaxation times. 

The complex electric permittivity data can be fitted using the modified Cole-Cole equation as given below \cite{Cole_JCP_41, Ang_PRB_00, Sundarakannan_JAP_03, Raut_JAP_18}
\begin{equation}
    \epsilon^*=\epsilon^{'} - i \epsilon^{''}=\epsilon_{\infty} + \frac {\epsilon_0-\epsilon_{\infty}}{1+ (\iota \omega \tau)^{1-\alpha}}
\label{epsilon}
\end{equation}
Here $\epsilon_0$ and $\epsilon$$_\infty$ are the static and high-frequency values of electrical permittivity, respectively, $\omega$ is the applied angular frequency, $\tau$ is the mean relaxation time, and $\alpha$ is a parameter measures the distribution of relaxation times and represents the deviation from ideal Debye response. In an ideal Debye-type relaxation process there is no interaction between dipoles ($\alpha$=0) producing a sharp relaxation peak. If we consider significant interactions among the dipoles in this case we have $\alpha$$ >0$, which shows a broad relaxation peak with a distribution of relaxation times \cite{Ang_PRB_00, Raut_JAP_18}. Using the equation \ref{epsilon}, the real ($\epsilon$$^{'}$) and imaginary ($\epsilon$$^{''}$) parts of electric permittivity are expressed as below \cite{Cole_JCP_41,Ang_PRB_00, Sundarakannan_JAP_03, Raut_JAP_18} 
\begin{subequations}
\begin{equation}
\epsilon^{'}= \epsilon_{\infty} + \frac{\Delta \epsilon}{2} [1-\frac{sinh(\beta z)}{cosh(\beta z)+ sin (\frac{\beta \pi}{2})}]
\label{E'}
 \end{equation}  
\begin{equation}
\epsilon^{''}=  \frac{\Delta \epsilon}{2} [\frac{sin (\frac{\beta \pi}{2})}{cosh(\beta z)+ cos (\frac{\beta \pi}{2})}] 
\label{E''}
 \end{equation}  
\end{subequations}
Here, z~=~ln($\omega$$\tau$), $\Delta$$\epsilon$=~$\epsilon$$_0$-~$\epsilon$$_\infty$ and $\beta$=1-$\alpha$. Using equation \ref{epsilon}, \ref{E'} and \ref{E''} the real ($\epsilon$ $^{'}$) and imaginary ($\epsilon$ $^{''}$) part of electric permittivity can be fitted and various parameters like $\epsilon$$_0$, $\epsilon$$_\infty$, $\beta$ and $\tau$ can be obtained as a function of temperature. The imaginary part of electric permittivity, as presented in Figure~\ref{IE}, shows the broad relaxation peak indicates non-Debye type relaxation behavior for all the Ni doped samples.

\subsection{Impedance Spectroscopy:}

\begin{figure*} 
\centering
\includegraphics[width=1.0\textwidth]{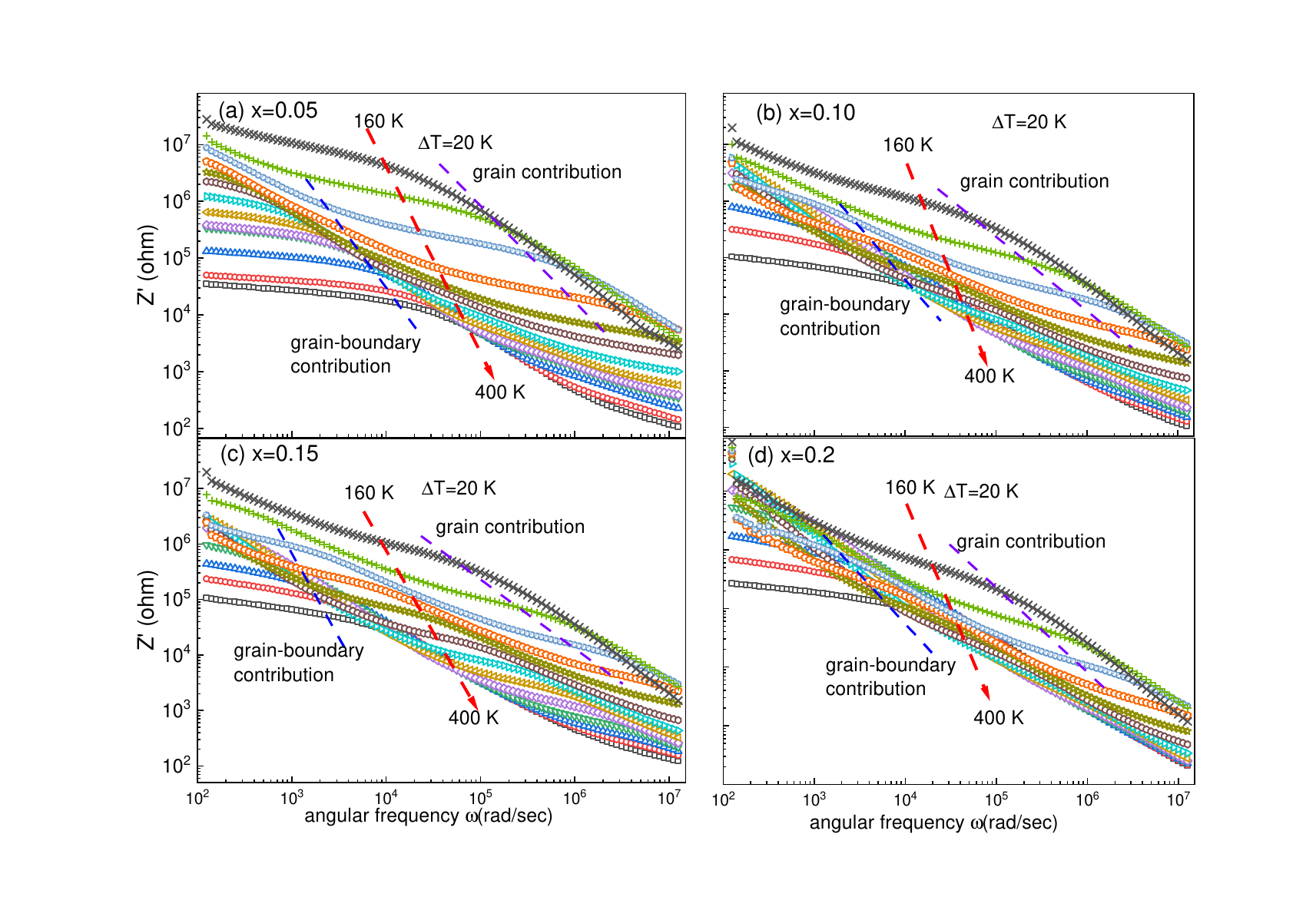}
\caption {The frequency dependence of real impedance data for the Na$_{3+2x}$Zr$_{2-x}$Ni$_{x}$Si$_2$PO$_{\rm 12}$ $(x=0.05-0.2)$ samples at various selected temperatures are shown in panels (a--d). The two arrows in these panels show the relaxation due to grain-boundary (at lower frequencies side) and grains (at higher frequency side). The arrow directions show an increase in relaxation towards high frequency side with an increase in temperature.} 
\label{Z'}
\end{figure*}

\begin{figure*}
    \centering
    \includegraphics[width=1.0\textwidth]{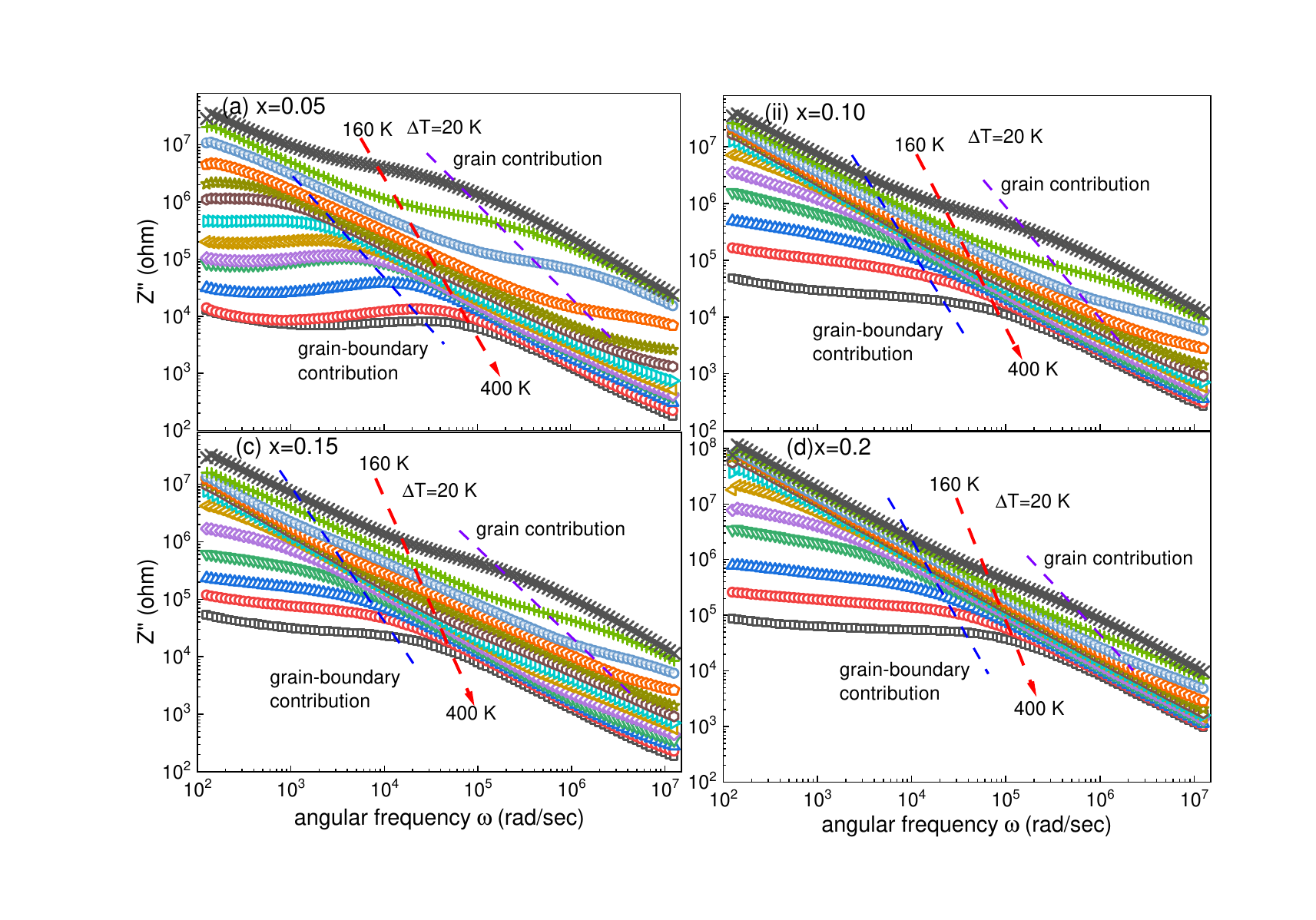}
    \caption{The imaginary part of total impedance for the Na$_{3+2x}$Zr$_{2-x}$Ni$_{x}$Si$_2$PO$_{\rm 12}$ $(x=0.05-0.2)$ samples at various temperatures as a function of frequency are shown in panels (a--d). The arrow shows an increase in relaxation peak towards the high-frequency side with an increase in temperature. All these curves deviate from the perfect circular arc representing the non-Debye type relaxation behavior.}
    \label{Z''}
\end{figure*}

The relaxation inside the material is due to grains, grain boundaries, electrode-interface contributions, or combinations of these. To find out the exact origin of the relaxation process, the combined analysis of impedance data and electric modulus are required \cite{Pu_JALCOM_16, Rehman_JAP_15}. Figures~\ref{Z'} and \ref{Z''} show the variation of real $(Z')$ and imaginary $(Z'')$ part of total impedance $(\lvert Z \rvert)$, respectively, as a function of frequency at various selected temperatures. All the samples have higher values of impedance at lower temperatures and frequencies, and the impedance decreases with further increase in the temperature and frequency. It is observed that both (real and imaginary) impedance spectra show strong frequency dispersions. The relaxation peak in impedance spectra is observed, once the applied frequency became equal to the hopping frequency of carriers. The imaginary part of total impedance $(Z'')$ shows two types of relaxation peaks (as indicated by the arrows) and these peaks at the lower frequency side are due to the grain-boundary relaxation and high frequency peaks correspond to the grain or bulk contribution in the total relaxation. More importantly, we find that both the relaxation peaks shift towards the higher frequency side with an increase in temperature for all the samples suggesting the thermally activated relaxation process. The grain peaks that appear at the high-frequency side of the spectra become outside of the measured frequency range at high temperatures. The values of real impedance $(Z')$ are found to decrease with an increase in frequency and temperature for all the samples, suggesting an increase in conductivity due to the possible release of space charge and reduction in the barrier height. The decrease in impedance with temperature is due to an increase in the mobility of charge carriers and a decrease in the density of trapped charge carriers  \cite{Lily_JALCOM_08, Taher_MRB_16,Idrees_JPD_10}. The broadening of the relaxation peak in impedance spectra shows a distribution of relaxation times. This relaxation phenomenon at lower temperatures appears due to immobile charge carriers, while defects and vacancies are responsible for high-temperature relaxation \cite{Nasari_CI_16}. The higher values of relaxation frequency at high temperatures suggest the temperature-dependent mobility of charge carriers over the measured temperature range \cite{Karmakar_JAP_20}. The relaxation peak in $(Z'')$ spectra shifts towards the high-temperature side showing the thermally activated relaxation. The asymmetric broadening shows the spread of relaxation time indicating the non-Debye type relaxation in the system for all samples. 

\subsection{Electric modulus study}
 
The complex electric modulus spectroscopy was introduced by Macedo for the analysis of dynamical electrical transport properties of non-conducting samples, as in modulus study the smallest capacitance exhibits the largest peak in the $M''$ vs frequency plots \cite{Macedo_PCG_1973}. The electric modulus study is used to find the charge carrier relaxation, ions hopping rate, and conduction mechanism via relaxation process in the electric field while electric displacement remains constant \cite{Moynihan_PCG_73, Macedo_PCG_1973}. Also, the analysis can provides information about the electric insulation, accumulation of charge carriers, carrier transportation, and polarization in the presence of electric field. These properties are useful for the multifunctionality of ceramic material as it provides the real cause of dielectric relaxation and it is applicable for conducting, and non-conducting as well as for ionic conducting materials. There is an advantage of representing the relaxation in modulus formalism as it suppresses the electrode polarization contribution \cite{Singh_JALCOM_17}. The electric modulus ($M$) and its real ($M'$) and imaginary ($M''$) parts are given by the following equation: 
\begin{subequations}
\begin{equation}
    \label{M}
    M^*(\omega)=M'(\omega)-jM''(\omega)=j \omega C_{0} Z^*=j \omega C_{0}(Z'-jZ'')
    \end{equation}
  \begin{equation}
    \label{M-a}
      M'(\omega)=\omega C_0 Z'' 
  \end{equation}
  \begin{equation}
    \label{M-b}
    M''(\omega)=\omega C_0 Z'
  \end{equation}
  \end{subequations}
Here, $\omega$ is the applied angular frequency, C$_0$ is the vacuum or geometrical capacitance given by (C$_0$= $\epsilon_0$$\frac{A}{d}$), where $d$ is the distance between the electrodes, A is the area of electrodes and $\epsilon_0$ is the permittivity of free space.

\begin{figure*} 
 \centering
\includegraphics[width=1.0\textwidth]{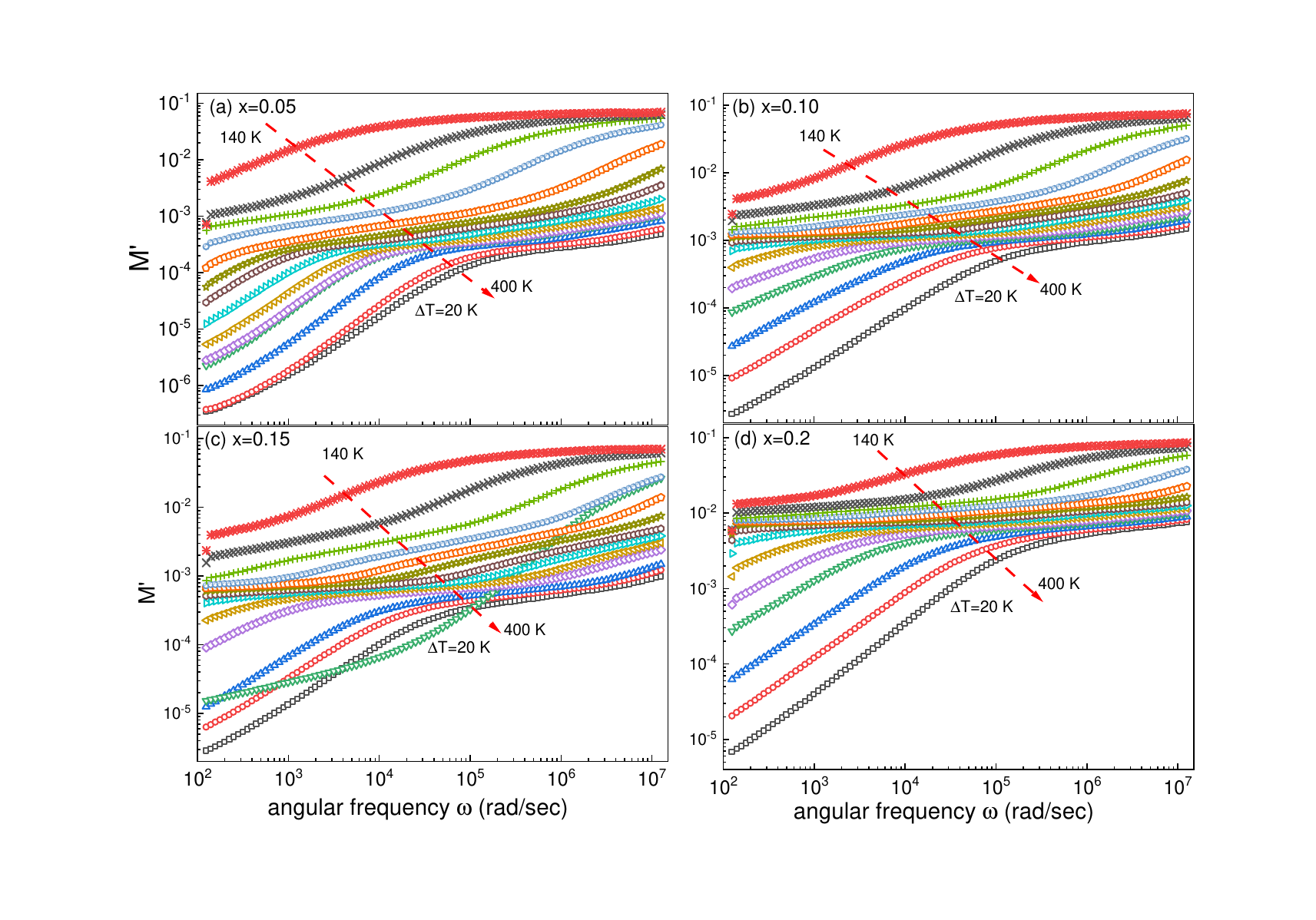}
\caption{The real part of electric modulus ($M'$) variation with frequency for the Na$_{3+2x}$Zr$_{2-x}$Ni$_{x}$Si$_2$PO$_{\rm 12}$ $(x=0.05-0.2)$ samples in the temperature range of 140--400~K are shown in panels (a--d). Here, the dashed arrow indicates the direction of increasing the temperature.} 
\label{M'}
\end{figure*}

\begin{figure*} 
 \centering
\includegraphics[width=1.0\textwidth]{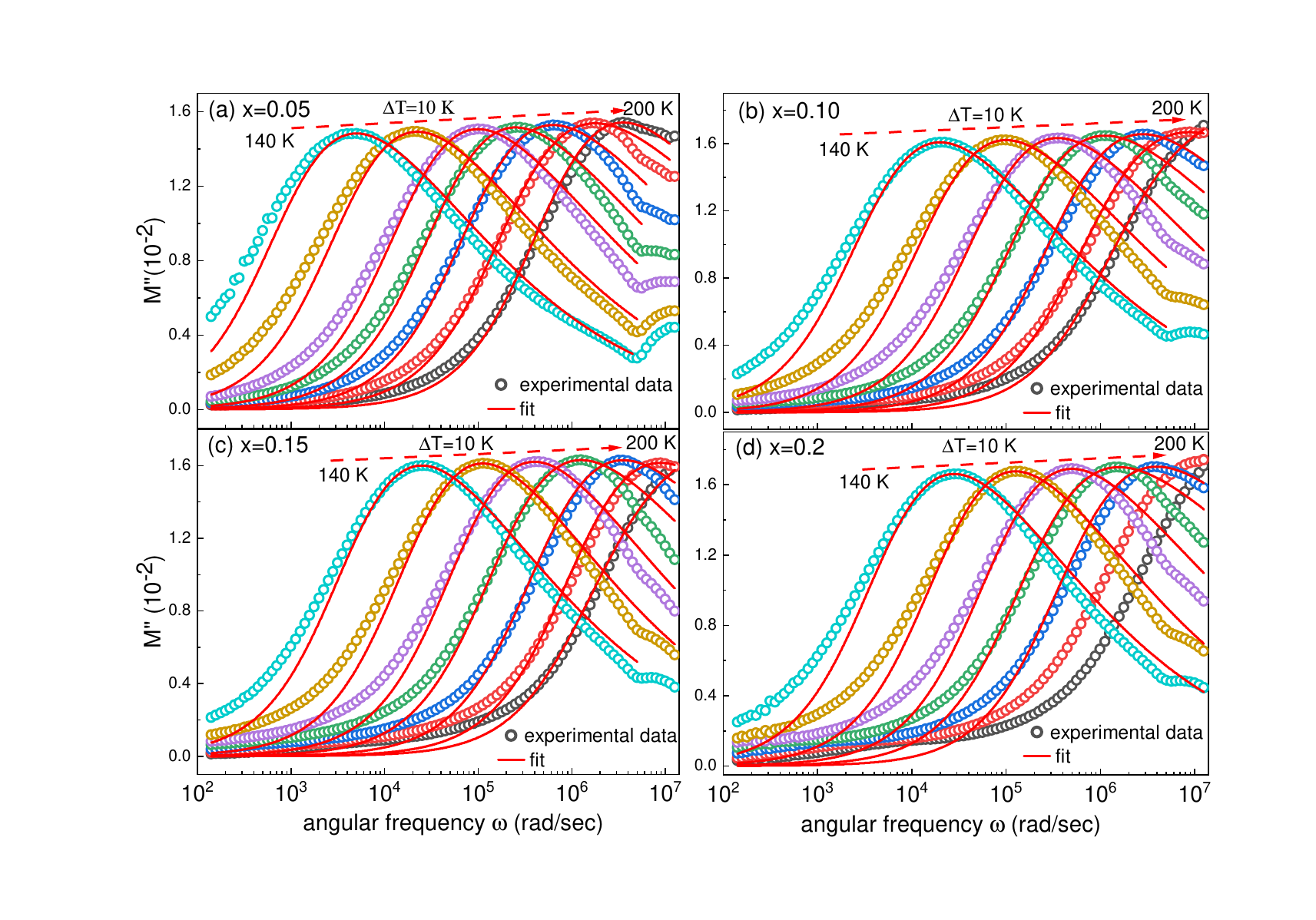}
\caption {The imaginary part of electric modulus $M''$ for the Na$_{3+2x}$Zr$_{2-x}$Ni$_{x}$Si$_2$PO$_{\rm 12}$ $(x=0.05-0.2)$ samples in the temperature range of 140--200~K are shown in panels (a--d). Here, open symbol represents the experimental data and the solid red lines represent the fit to the KWW function. The dashed black arrows in each panel indicate the direction of the relaxation peak shift with increasing the temperature. } 
\label{M''}
\end{figure*}

\begin{figure*} 
 \centering
\includegraphics[width=1.0\textwidth]{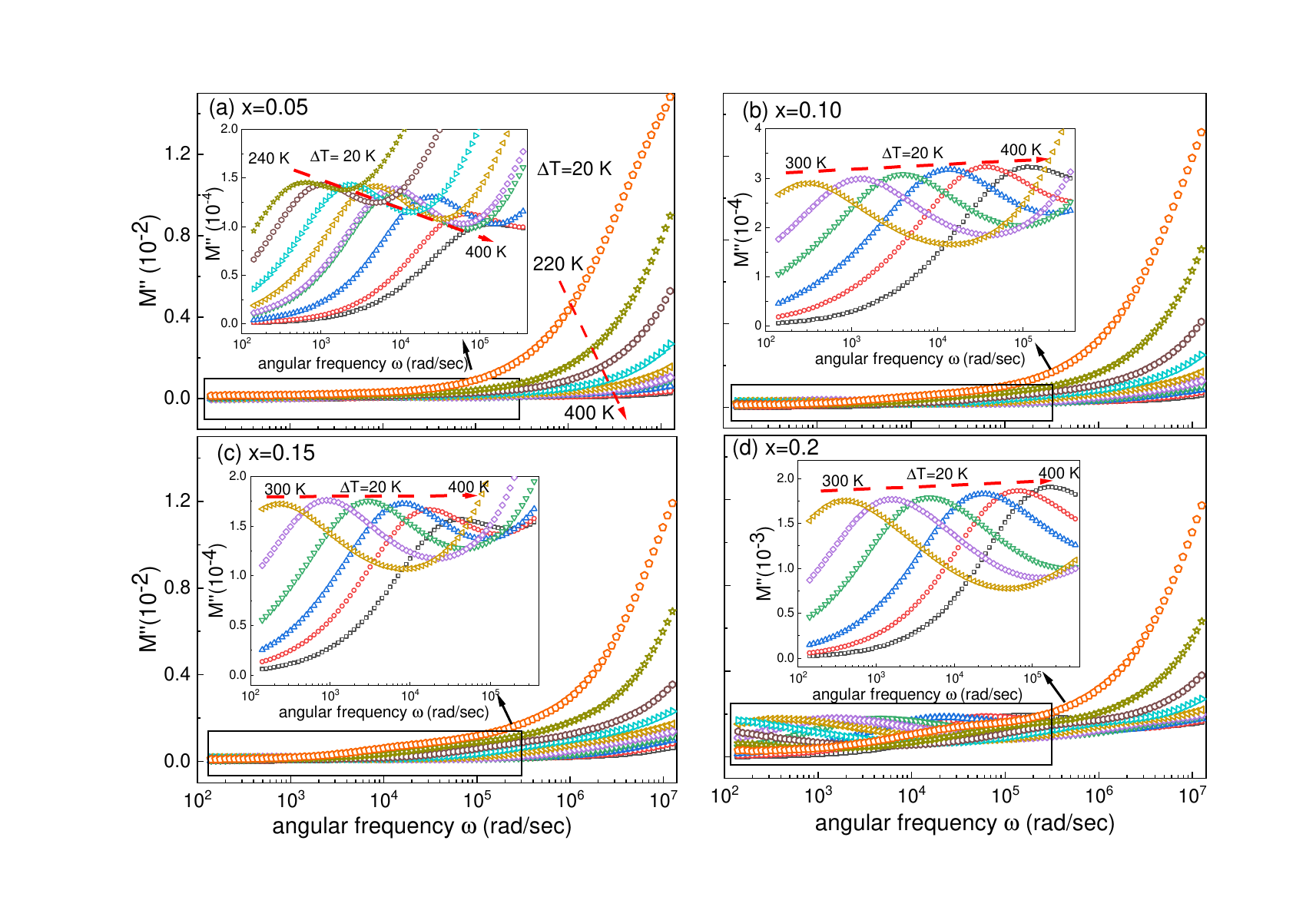}
\caption {The imaginary part of electric modulus $M''$ for the Na$_{3+2x}$Zr$_{2-x}$Ni$_{x}$Si$_2$PO$_{\rm 12}$ $(x=0.05-0.2)$ samples in the temperature range of 220--400~K are shown in panels (a--d). Here, the dashed arrows indicate the increase in relaxation frequency with an increase in the sample temperature.} 
\label{M''_1}
\end{figure*}

The frequency dependence of real ($M'$) part of complex electric modulus ($M^*$) for the Na$_{3+2x}$Zr$_{2-x}$Ni$_{x}$Si$_2$PO$_{\rm 12}$ $(x=0.05-0.2)$ samples are shown in Figures~\ref{M'}(a--d) at different temperatures. We find that the real part of electric modulus $M'$ approaches zero value at lower frequencies for all the samples (Figure~\ref{M'}) suggesting negligible effect of electrode polarization in the modulus formulation. The smaller values of $M'$ at lower frequencies are due to the lack of restoring forces governing the mobility of charge carriers under the influence of the applied electric field. At higher frequencies, a step-like behavior is observed in $M'$ spectra showing a tendency to approach towards $M_\infty$ (the asymptotic value of $M'(\omega)$), which confirms the capacitive nature of all the samples. The ($M'$) variation with frequency for all the samples shows the dispersed behavior due to conductivity relaxation and short-range mobility of charge carriers. It is found that the dispersion is high at higher temperatures due to lower restoring forces giving an increase in the {\it a.c.} conductivity under the influence of the applied electric field with temperature. The magnitude of $M'$ decreases with an increase in temperature for all the samples attributed to the temperature-dependent relaxation \cite{Singh_JALCOM_17, Sondarva_JALCOM_21, Moynihan_PCG_73}. 

\begin{figure} 
\centering
\includegraphics[width=0.48\textwidth]{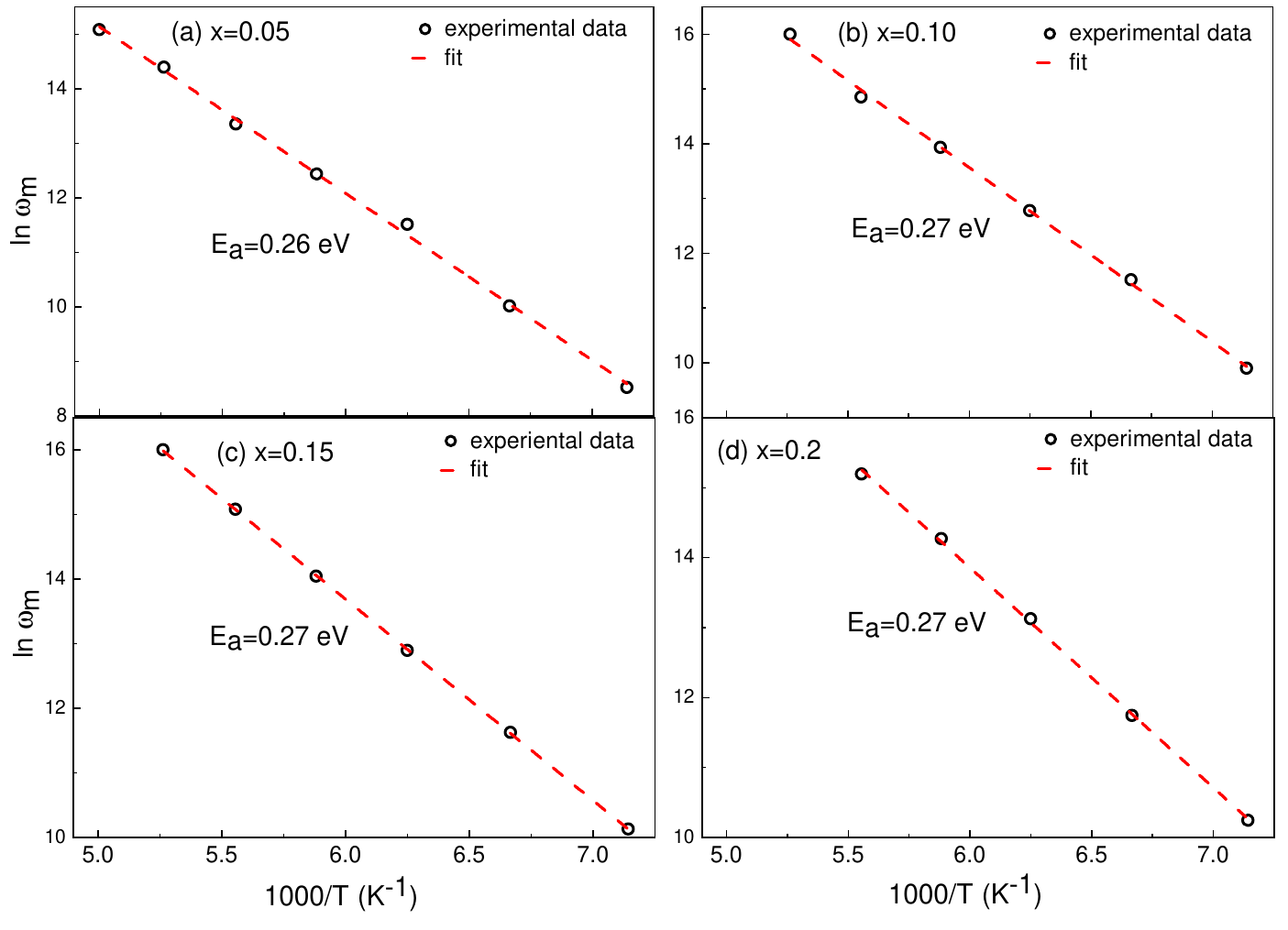}
\caption {The Arrhenius activation energy of relaxation for the Na$_{3+2x}$Zr$_{2-x}$Ni$_{x}$Si$_2$PO$_{\rm 12}$ $(x=0.05-0.2)$ samples, shown in panels (a--d) at room temperature. Here, the open symbols represent the experimental data and the dashed lines represent the linear fit to the equation 7.} 
\label{ME}
\end{figure}

The imaginary part of electric modulus ($M''$) has also been studied in the temperature range of 140--400~K and the data are shown in Figures~\ref{M''} and \ref{M''_1}, respectively for the Na$_{3+2x}$Zr$_{2-x}$Ni$_{x}$Si$_2$PO$_{\rm 12}$ $(x=0.05-0.2)$ samples. At lower temperatures, shown in Figure~\ref{M''}, the grain contributions are dominating in the relaxation while at high-temperature side grain-boundary contributions are dominating in the total relaxation process. The imaginary modulus spectra ($M''$) show the asymmetric nature of the relaxation peak and the peak frequency $\omega_m$ (the frequency at which $M''$ is maximum) shifts towards the high-frequency side with an increase in temperature, suggesting the ionic nature having temperature dependent relaxation for all the doped samples. The shifting in relaxation peak towards the higher frequency side is due to free charge accumulation at the interface. This charge accumulation increases with an increase in temperature giving the relaxation occurs at higher frequency. The increased relaxation frequency leads to decreased relaxation time due to enhanced charge carrier mobility. The heights of these relaxation peaks are nearly constant with temperature suggesting the associated capacitance is weakly temperature dependent \cite{Ortega_PRB_08, Liu_PRB_04}. The intensity of grain peaks (dominating at lower temperatures) is 10$^2$ times larger than the grain-boundary peaks (dominating at high temperatures) showing the grain-boundary capacitance is smaller than grain capacitance. The increase in relaxation rate (decreased relaxation time) with temperature is due to the thermal activation of charge carriers, as the thermal energy increases the mobility of charge carriers increases leading to a decrease in relaxation time \cite{Singh_JALCOM_17, Taher_MRB_16}. In the frequency range below the relaxation peak, the charge carriers move freely and drift over a longer distance, and above the $M''$ peak, the carriers are confined over a potential wall and move inside the walls having localized motions only. This means the position of the relaxation peak represents the transition from the long-range (below relaxation peak) to the short-range (above relaxation peak) ordering of charge carriers \cite{Sinha_PC_15}. The position of $\omega_{max}$ (position corresponds to the $M''_{max}$) shifts towards the higher frequency side with an increase in temperature suggesting that it is a thermally activated process. It is also found the peak frequency $\omega_m$ increased towards the high-frequency side with an increase of Ni doping due to an increased number of charge carriers. These carriers lead to a decrease in relaxation time, hence an increase in relaxation frequency. The relaxation time ($\tau$) is obtained using the relation $\omega_m$$\tau_m$=1. The activation energy of relaxation time can be calculated using the Arrhenius law of thermal activation given by: 
\begin{equation}
    \label{omega-Arr}
    \omega_{m} = \omega_0 ~ exp (\frac{-E_a}{k_BT})
\end{equation}
Here, $\omega_0$ is the pre-exponential factor, k$_B$ is the Boltzmann constant, T is the absolute measured temperature and $E_a$ is the activation energy of the relaxation. The activation energy is calculated from the linear least square fit of ln$(\omega_{m})$ versus 1000/T data, which found to be nearly the same for all the samples and the value is around 0.27~eV, as shown in Figure~\ref{ME}. This represents the similar type of relaxation process in all the samples and the magnitude of activation energy suggests the transport phenomena are due to the hopping mechanism. The activation energies for high-temperature relaxation are calculated using Figure~\ref{M''_1} (shown in Figure 1 of supplementary Information \cite{SI}), where the larger values indicate that higher activation energy is needed for the grain boundary conduction as compared to the grains. It is also observed in the imaginary modulus ($M''$) spectra that the relaxation peaks are broader as compared to the ideal Debye response. The asymmetric nature and broadening of $M''$ peaks show the non-exponential behavior of conductivity relaxation. In this type of behavior, the ions migrate via hopping by time-dependent mobility in the vicinity of other charge carriers \cite{Nasri_CerInt_16}. 

As we know that the imaginary part of the electric modulus ($M''$) represents the energy loss inside the material under the influence of the applied electric field. So, the $M''$ can be expressed in terms of relaxation function as given below: 
\begin{equation}
M^*= M_\infty [1-\int_{0}^{\infty}exp(-\omega t) \frac{d\phi}{dt} dt]  
\end{equation}
here, $\phi (t)$ represents the evolution of the electric field inside the material, which can be expressed as Kohlraush–Williams–Watts (KWW) decay function:  
\begin{equation}
\phi (t)=[\exp{(-\frac{t}{\tau})^\beta}],
\end{equation}
where, $\tau$ is the characteristic relaxation time of charge carriers and $\beta$ is the Kohlrausch parameter known as stretched exponent related to the asymmetry observed in the relaxation peak, which varies between 0 to 1 and characterizes the type of relaxation behavior (Debye or non-Debye) as well as represents the interaction among charge carriers. For the ideal Debye type relaxation where dipole-dipole interactions are neglected and the value of $\beta$ is found to be one. However, the value of $\beta$ deviates from unity when considering the significant interaction among the dipoles for non-Debye type behavior. To analyze the imaginary modulus data directly in the frequency domain, Bergman has modified the KWW function, as given below \cite{Bergman_JAP_00, Taher_MRB_16, Karmakar_JPCM_19}: 
\begin{equation}
\label{eq-M}
    M''=\frac{M''_{max}}{[(1-\beta)+(\frac{\beta}{1+\beta})[\beta(\frac{\omega_m}{\omega})+(\frac{\omega}{\omega_m})^\beta]}
\end{equation}
where, $M''_{max}$ is the peak maximum in the $M''$ spectra, $\omega_{m}$ is the frequency corresponding to the peak maximum and $\beta$ is the stretched exponent that decides the range of relaxation time distributions or type of interaction among the charge carriers. The value of $\beta$ is calculated using equation \ref{eq-M} and the obtained values are in the range from 0.28--0.41 (less than unity), which confirm the non-Debye type relaxation among the carriers in all the Ni-doped NASICON samples. The value of $\beta$ is found to be  increased with temperature suggesting the weak interaction among charge carriers in the material. 

\begin{figure*} 
\centering
\includegraphics[width=1.0\textwidth]{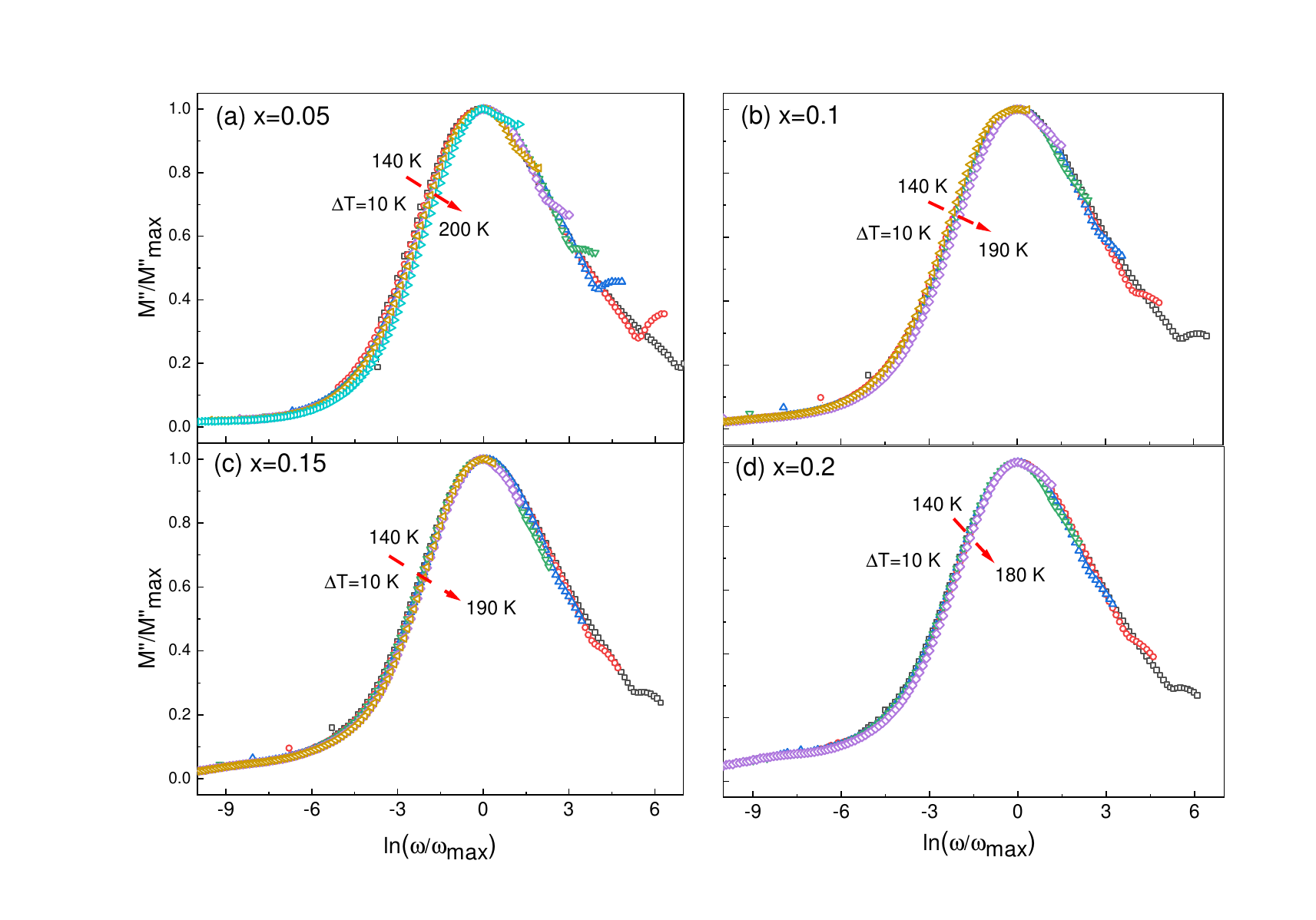}
\caption{The scaling analysis of imaginary electric modulus of Na$_{3+2x}$Zr$_{2-x}$Ni$_{x}$Si$_2$PO$_{\rm 12}$ $(x=0.05-0.2)$ samples at various temperatures are shown in figure (a-d). In all these figures the open symbol represents the experimental data and the arrow shows the increase in temperature. The merging of all these curves into a single master curve shows a similar type of relaxation.} 
\label{MS}
\end{figure*}

Further, in order to determine the type of relaxation process, we scaled the imaginary part of electric modulus $(M'')$ for all the Ni-doped samples. The scaling at low temperatures is shown in Figure~\ref{MS} where all the curves merge with each other for all the samples. The overlapping of scaled modulus spectra in a single master curve at low temperatures indicates the relaxation at low-temperature regions by a similar mechanism. The scaling at higher temperatures shows the merging of the curves below the peak frequency and dispersed variation above the peak frequency (see Figure 2 of Supplementary Information \cite{SI}). The scaling analysis indicates the relaxation process is independent of temperature in the low-temperature region and varies in the high-temperature range \cite{Kumar_MCP_13, Baskaran_JAP_02, Singh_JALCOM_17}. The type of conduction mechanism (long-range hopping or short-range hopping) followed by charge carriers can be determined using the combined plot of imaginary impedance ($Z''$) with electrical modulus ($M''$). It has been reported that if the peaks in imaginary impedance and electric modulus overlap with each other, the relaxation is due to long-range mobility, whereas, if these curves do not coincide with each other the relaxation is due to the short-range mobility of carriers \cite{Ksentini_APA_20}. The mismatch between these spectra is due to multiple relaxations and changes in the polarization mechanism. Figure 3 in Supplementary Information \cite{SI} shows the variation of $Z''$ and $M''$ with frequency at 300~K. The amplitude of impedance is proportional to the resistive element, while in the case of electric modulus, it depends on the lowest capacitive element. The peaks in these curves do not overlap, suggesting the short-range mobility of charge carriers and non-Debye type relaxation in all the studied samples \cite{Sondarva_JALCOM_21, Raut_JAP_18, Ksentini_APA_20}.      

\begin{figure*} 
\centering
\includegraphics[width=1.0\textwidth]{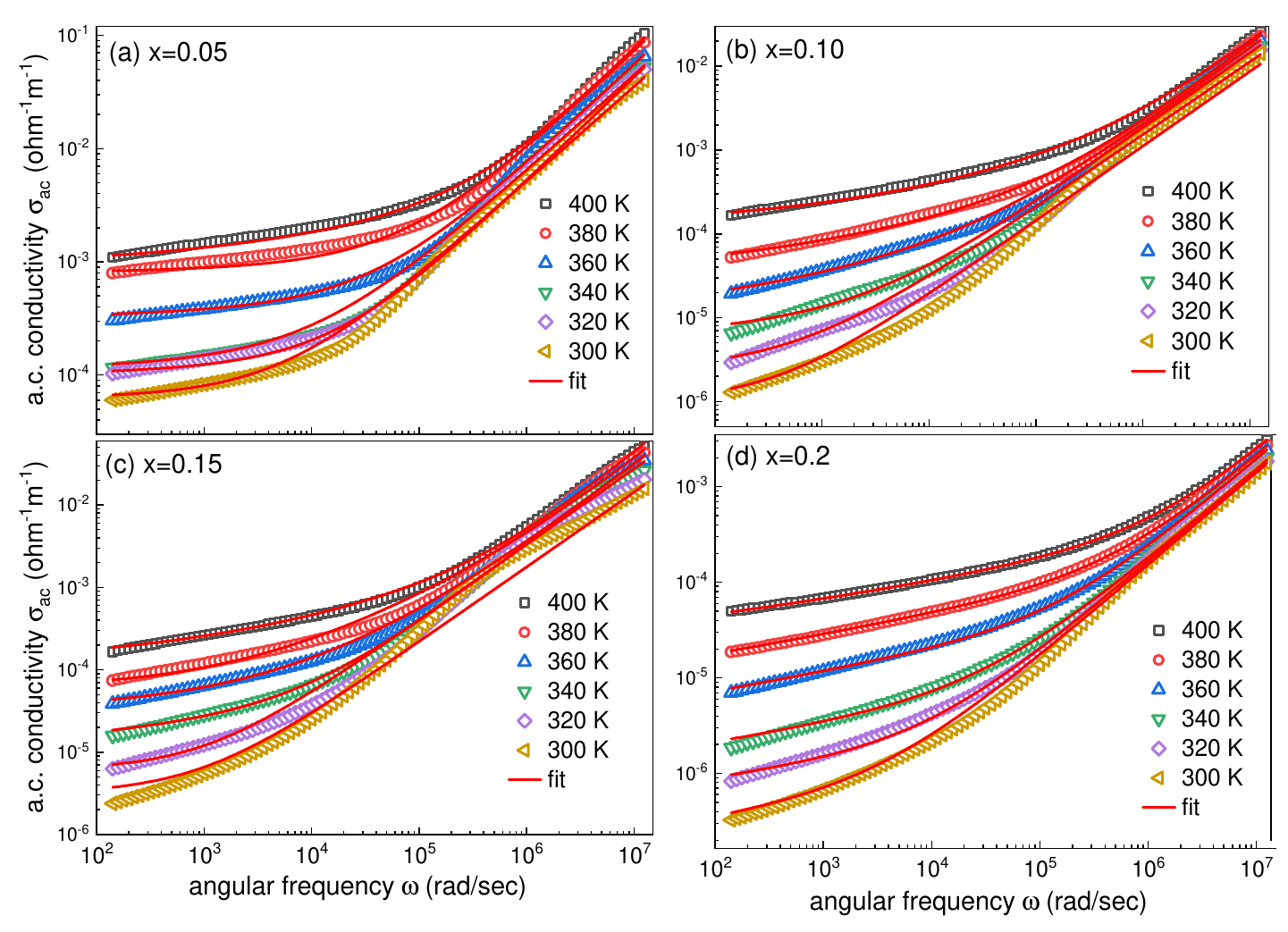}
\caption{The {\it a.c.} conductivity variation of Na$_{3+2x}$Zr$_{2-x}$Ni$_{x}$Si$_2$PO$_{\rm 12}$ $(x=0.05-0.2)$ samples at various temperatures are shown in figure (a-d). In all the figures the open symbol represents the experimental data and the solid line represents the fit using the double power law equation 13.} 
\label{acs}
\end{figure*}

\subsection{{\it a.c.} conductivity}

In order to understand the {\it a.c.} conductivity of Ni-doped NASICON samples Na$_{3+2x}$Zr$_{2-x}$Ni$_{x}$Si$_2$PO$_{\rm 12}$ $(x=0.05-0.2)$, we perform the measurements in the frequency and temperature range of 20 Hz-2 MHz and 300--400~K,  as shown in Figure~\ref{acs}. Interestingly, the conductivity is found to increase with temperature and frequency for all the samples. The {\it a.c.} conductivity variation with temperature shows the dispersed behavior at lower frequencies and merging behavior at higher frequencies. Moreover, we find that the {\it a.c.} conductivity decreases when the Ni concentration increases as the the scattering among the carriers enhanced with doping. The {\it a.c.} conductivity data are further analyzed in order to determine the type of conduction mechanism over the measured frequency and temperature range. The contribution in {\it a.c.} conductivity arises from the deep electronic states, which are in close vicinity of the Fermi level, and found to be a combined response of relaxation, localized charge carriers hopping, and diffusion of charge carriers. These charge carriers could migrate over short-range or long-range under the influence of the applied electric field. Also, the {\it a.c.} conductivity depends upon the geometry of the sample, which can be calculated at a particular temperature using the following relation \cite{Raut_JAP_18}:  
\begin{equation}
\sigma_{ac}(\omega)= \frac{d}{A} \left[ \frac{Z'}{Z'^2+Z''^2} \right]
\end{equation}
Here, $d$ is the thickness of the dielectric material, $A$ is the area of the capacitor electrode, $Z'$ and $Z''$ are the real and imaginary parts of the total impedance $Z$. 

The frequency-dependent behavior of {\it a.c.} conductivity can be explained using the jump relaxation model given by Funke \cite{Funke_PSSC_93, Sumi_Jap_10, Kahouli_JPCA_12}. According to this model, the charge carriers undergo successful hopping for longer periods at lower frequencies, which gives the long-range translational motion of carriers and provides the frequency-independent conductivity. While at higher frequencies, two competing processes, unsuccessful and successful hopping occur due to carriers' shorter periods, and  their ratio decides the abrupt behavior in conductivity at a higher frequency. The conventional analysis of {\it a.c.} conductivity with frequency for various types of materials follows the Jonsher power law as given by \cite{Raut_JAP_18, Nasari_CI_16, Jonscher_Nature_77,  Sumi_Jap_10, Kahouli_JPCA_12}:   
\begin{equation} 
\label{power-law}
	\sigma_{(ac)} (\omega, T) = \sigma_{dc} (T) +A ~\omega^n
	\end{equation}
here, $\sigma_{dc}$(T) is the temperature-dependent {\it d.c.} contribution in total conductivity, coefficient $A$, and exponent $n$ are temperature-dependent and depends on the type of materials. The second term in equation~\ref{power-law} related to the dispersive behavior of total conductivity. However, none of the Ni-doped samples follow the power law relation within the measured temperature range. The frequency-dependent {\it a.c.} conductivity shows the two types of dispersed behavior (i) at low frequency, the dispersed region is due to the contributions from grain boundary (having the large capacitance value), and (ii) at higher frequency abrupt behavior is due to contributions from the grains (having the lower capacitance value). To describe the different contributions in conductivity, we apply the jump relaxation model and fitted the data with the double power law given by \cite{Ortega_PRB_08, Sumi_Jap_10, Rao_JSNM_20, Moualhi_JALCOM_22, Chen_JAP_10}:   
\begin{equation} 
\label{ double power-law}
	\sigma_{(ac)} (\omega, T) = \sigma_{dc} (T) +A ~\omega^m +B ~\omega^n
	\end{equation}
here, the $1^{st}$ term represents the long-range translational hopping of charge carriers due to the availability of large time scale, which gives the direct current contribution in the total conductivity, and coefficients $A$, $B$, and exponents $m$, $n$ are material and temperature dependent parameters. The $2^{nd}$ term (A$w^m$) belongs to the middle frequency (large capacitance) region of the conductivity, which corresponds to the translational motion due to short-range hopping over the potential barriers and gives the grain-boundary contribution, where the exponent $m$ is having the values between 0 and 1. Whereas, the $3^{rd}$ term (B$w^n$) at higher frequencies (smaller capacitance) corresponds to the localized or reorientation hopping motion within the grain, where the exponent $n$ values lie between 0 and 2. 

The {\it a.c.} conductivity data fitted using the double power law are shown in Figure~\ref{acs}. The variation of the fitted exponent $m$, $n$ are related to the contributions from grain boundary, and grains to the total {\it a.c.} conductivity in the measured temperature. It is found from the fitted data that the values of $m$ are varied between 0.81 to 1 and $n$ varied between 0.1 to 0.5. The variation in the values of $m$ and $n$ exponents give useful information about the type of conduction mechanism with temperature, i.e., for small polaron hopping the exponent increases with temperature while for large polaron hopping, exponent values decrease with temperature \cite{Ortega_PRB_08, Raut_JAP_18}. It is found that the values of $m$ and $n$ increase with temperature showing the small polaron hopping type of conduction in our sample over the measured frequency range (shown in Figure 4 of Supplementary Information \cite{SI}). Figure~\ref{acs} shows that the dispersive region is shifted towards the high-frequency side with an increase in temperature, which can be explained as follows: at higher applied frequencies, the charge carriers have less time scale to switch their direction with applied field and result in unscusseful hopping of carriers to their neighboring sites. As the temperature is increased the average thermal energy available with charge carriers increases resulting in increased lattice vibrations. These vibrations help the carriers to have the long-range translational hopping even for the short period available at higher frequencies. This long-range translational hopping provides the extension of the dispersive region toward higher frequency with temperature for all the samples \cite{Sumi_Jap_10, Rao_JSNM_20, Moualhi_JALCOM_22, Chen_JAP_10}.    

\section{\noindent ~Summary and Conclusions}
 
In summary, we successfully prepared the Ni-doped samples having the chemical composition of Na$_{3+2x}$Zr$_{2-x}$Ni$_{x}$Si$_2$PO$_{\rm 12}$ $(x=0.05-0.2)$ using the solid-state reaction, which show the monoclinic phase with C 2/c space group along the small amount of ZrO$_2$ and Na$_3$PO$_4$ impurity phases. The scanning electron microscope images show the non-uniform distribution of grains with dense microstructure. The {\it d.c.} resistivity measurements show the semiconducting nature and follow the Arrhenius-type thermal conduction. The temperature and frequency dependence of the dielectric constant were explained using the space-charge polarization and Maxwell-Wagner relaxation mechanism. The relaxation peak in dielectric loss data follows the Arrhenius-type relaxation mechanism with nearly the same activation energy for all the samples. The real and imaginary parts of the dielectric constant shows the broad distribution of relaxation peak indicating the non-Debye type relaxation. The electric impedance analysis with frequency at various temperatures shows two types of relaxation peaks corresponding to the grain and grain-boundary relaxation. The peak frequency of grain and grain-boundary relaxation shifts towards the higher frequency side with an increase in temperature. The imaginary part of the electric modulus is fitted using the KWW function confirming the non-Debye type relaxation. The electric modulus peak frequency follows the Arrhenius type relaxation with similar type relaxation energy at lower temperatures. The scaling analysis shows a similar type of relaxation at lower temperatures, while different types of relaxation at high temperatures above the peak frequency. The {\it a.c.} conductivity data were fitted using the double power law confirming the contribution from grain and grain-boundary conductivity. The temperature dependence of the double power law exponent shows the small polaron hopping type conduction over the measured temperature range for all the samples.  

\section*{\noindent ~Supplementary Material}

See the supplementary information for further analysis figures of activation energy, electric modulus, and Power law exponents.

\section{\noindent ~Acknowledgments}

RM  thanks IUAC for providing the experimental facilities. The authors are thankful to IUAC for extending the FE-SEM facility funded by Ministry of Earth Sciences (MoES) under the Geochronology project. RSD acknowledges the financial support from SERB-DST through a core research grant (project reference no. CRG/2020/003436). 

\section*{\noindent ~Declaration of competing interest}

The authors declare that they have no known competing financial interests or personal relationships that could have appeared to influence the work reported in this paper.

\section*{\noindent~Data Availability}

The data that support the findings of this study are available from the corresponding author upon reasonable request.

\end{document}